\documentclass[12pt]{article}
\pdfoutput=1
\usepackage{jheppub}
\usepackage[T1]{fontenc}
\usepackage{color}
\usepackage{float}
\usepackage{url}
\usepackage{braket}
\usepackage{amsmath}
\usepackage{amsthm}
\usepackage{amssymb}
\usepackage{amsfonts}
\usepackage{graphicx}
\usepackage{caption}
\usepackage{subcaption}
\usepackage{multirow,array}
\usepackage{qcircuit}
\usepackage{hyperref}
\usepackage{comment}
\definecolor{red}{rgb}{1,0.,0}

\subheader{\begin{flushright}
\texttt{UTWI-42-2023, IFT-UAM/CSIC-23-148}
\end{flushright}}
\title{Shock waves, black hole interiors and holographic RG flows}
\author{Elena Cáceres$^{a}$, Ayan K.~Patra$^{b}$ and Juan F. Pedraza$^{b}$}
\affiliation{$^a$Theory Group, Department of Physics, University of Texas, Austin, TX 78712, USA}
\affiliation{$^b$Instituto de Física Teórica UAM/CSIC, Calle Nicolas Cabrera 13-15, Madrid 28049, Spain}
\emailAdd{elenac@utexas.edu, a.patra@csic.es, j.pedraza@csic.es}

\abstract{We study holographic renormalization group (RG) flows perturbed by a shock wave in dimensions $d\geq 2$. The flows are obtained by deforming a holographic conformal field theory with a relevant operator, altering the interior geometry from AdS-Schwarzschild to a more general Kasner universe near the spacelike singularity. We introduce null matter in the form of a shock wave into this geometry and scrutinize its impact on the near-horizon and interior dynamics of the black hole. Using out-of-time-order correlators, we find that the scrambling time
increases as we increase the strength of the deformation, whereas the butterfly velocity 
displays a non-monotonic behavior. We examine other observables that are more sensitive to the black hole interior, such as the thermal $a$-function and the entanglement velocity.
Notably, the $a$-function experiences a discontinuous jump across the shock wave, signaling an instantaneous loss of degrees of freedom due to the infalling matter. This jump is interpreted as a `cosmological time skip' which arises from an infinitely boosted length contraction. The entanglement velocity exhibits similar dependence to the butterfly velocity as we vary the strength of the deformation. Lastly, we extend our analyses to a model where the interior geometry undergoes an infinite sequence of bouncing Kasner epochs.}

\begin{document}
\maketitle
\vspace{-0.2cm}

\section{Introduction}
Holographic duality holds considerable promise in unraveling the mysteries of the black hole interior.  Various avenues for exploring this enigmatic region through holographic duality include delving into analytically continued correlation functions \cite{Fidkowski:2003nf, Festuccia:2005pi,Horowitz:2023ury}, entanglement entropy \cite{Hartman:2013qma,Penington:2019npb,Almheiri:2019hni}, and computational complexity \cite{Stanford:2014jda,Brown:2015bva,Couch:2016exn,Belin:2021bga}. Among the most striking features within the black hole interior are the inevitable spacetime singularities. Spacelike singularities, a particular class of singularities where time seemingly `comes to an end,' pose distinct conceptual challenges. Moreover, they underscore profound similarities with cosmological solutions to Einstein's equations, e.g. those featuring big-bang or big-crunch singularities. Numerous endeavors have been undertaken to understand these singularities through the lens of holography \cite{Fidkowski:2003nf, Festuccia:2005pi,Horowitz:2023ury,Festuccia:2006sa, Grinberg:2020fdj,Pedraza:2021fgp,Leutheusser:2021frk,Rodriguez-Gomez:2021pfh,deBoer:2022zps}. 

One of the extensively examined black hole interiors in holography is that of the eternal Schwarzschild-AdS black hole, crucial for characterizing the thermofield double (TFD) state of the dual conformal field theory (CFT) \cite{Maldacena:2001kr}. While the exterior geometry of these black holes is dynamically stable, it is well known that their interior is not. It has been established that matter fields exhibit infinite growth as they approach a spacelike singularity, triggering substantial backreaction \cite{Belinsky:1970ew,Belinskii:1973zz,Belinskii:1981vdw}. Broadly speaking, the Schwarzschild singularity is precisely fine-tuned within the spectrum of potential late-time behaviors of gravity, rendering it an atypical late-time solution. Consequently, this inherent instability of the Schwarzschild singularity necessitates careful consideration in any holographic exploration of the black hole interior.

On general grounds, we expect typical black hole interiors to be inhomogeneous and anisotropic. Even so, if we limit ourselves to geometries that retain the spacetime symmetries of Schwarzschild-AdS, the Schwarzschild singularity is still highly fine-tuned. Motivated by these conceptual challenges, the authors of \cite{Frenkel:2020ysx} studied a class of black holes that result from deforming the dual theory with a relevant operator and found the emergence of a more general Kasner singularity as an endpoint of the interior's evolution.\footnote{See also \cite{Hartnoll:2020rwq, Hartnoll:2020fhc,Sword:2021pfm, Sword:2022oyg,Wang:2020nkd, Caceres:2021fuw, Mansoori:2021wxf, Liu:2021hap, Caputa:2021pad, Bhattacharya:2021nqj, Das:2021vjf,Caceres:2022smh,An:2022lvo,Auzzi:2022bfd, Liu:2022rsy,Mirjalali:2022wrg,Caceres:2023zhl,Blacker:2023ezy, Gao:2023rqc} for more recent developments on Kasner interiors in the context of holography.} These are precisely the type of singularities discovered by Belinsky-Khalatnikov-Lifshitz (BKL) in the early 70's \cite{Belinsky:1970ew,Belinskii:1973zz,Belinskii:1981vdw}. Our main objective is to understand this generic class of geometries when additional null matter is thrown into the black hole. The resulting shock wave geometries are dual to `quenched' states in the boundary CFT, which can be obtained by turning on an instantaneous perturbation at a given boundary time. Indeed, understanding black hole geometries with shock waves can provide a scope to analyze the nature of the spacetime singularity \cite{Horowitz:2023ury}. Particularly, by examining specific observables we may uncover novel signatures of the singularity and gain insights into how it is encoded in the boundary CFT.

This paper is structured as follows. In Section \ref{RG Kasner}, we begin with a brief overview of holographic RG flows featuring Kasner interiors. Subsequently, in Section \ref{Rg with a shock}, we study the injection of null matter into the deformed background, representing a quenched state in the dual theory. Holographically, the abrupt injection of matter manifests as a shock wave on top of the black hole, sent from one of the boundaries and reaching the singularity. We characterize the imprints of the shock wave on several field theory observables, such as four-point out-of-time order correlators (OTOC), computed in the heavy-heavy-light-light limit, the thermal $a$-function, and the entanglement velocity. Notably, in terms of the interior's evolution, the shock wave is shown to induce a `time skip' which we understand as a purely relativistic effect. In Section \ref{Bounces} we explore a scenario wherein the interior of the black hole exhibits an infinite sequence of bouncing Kasner epochs, akin to Misner’s mixmaster universe \cite{Misner:1969hg}. In this case, depending on the energy of the shock wave, the cosmological evolution may abruptly transition between epochs or even skip one or multiple epochs altogether. Finally, in Section \ref{discussion} we discuss our main conclusions and provide some interesting directions for future work.


\section{Kasner interiors and holographic RG flows: a review}{\label{RG Kasner}}
Let us start this section by reviewing the holographic RG flows proposed in \cite{Frenkel:2020ysx} and generalized to arbitrary dimensions in \cite{Caceres:2021fuw}. We take $(d+1)$-dimensional Einstein gravity ($d \geq 2$) with negative cosmological constant $\Lambda = -d(d-1)/2$ (setting the AdS radius to unity $L=1$), coupled to a scalar field $\phi$ with potential $V(\phi)$. We consider the minimal case of a free massive scalar field, in which the potential is $V(\phi) =\frac{1}{2}m^2 \phi^2$. The corresponding action is given by,
\begin{equation}
I = \frac{1}{16\pi G_{d+1}}\int d^{d+1}x \sqrt{-g}\left(R+d(d-1)-\frac{1}{2}\nabla^\alpha\phi \nabla_\alpha\phi-V(\phi)\right)\,.\label{action}
\end{equation}

\begin{figure}
    \centering
    \includegraphics[scale=0.15]{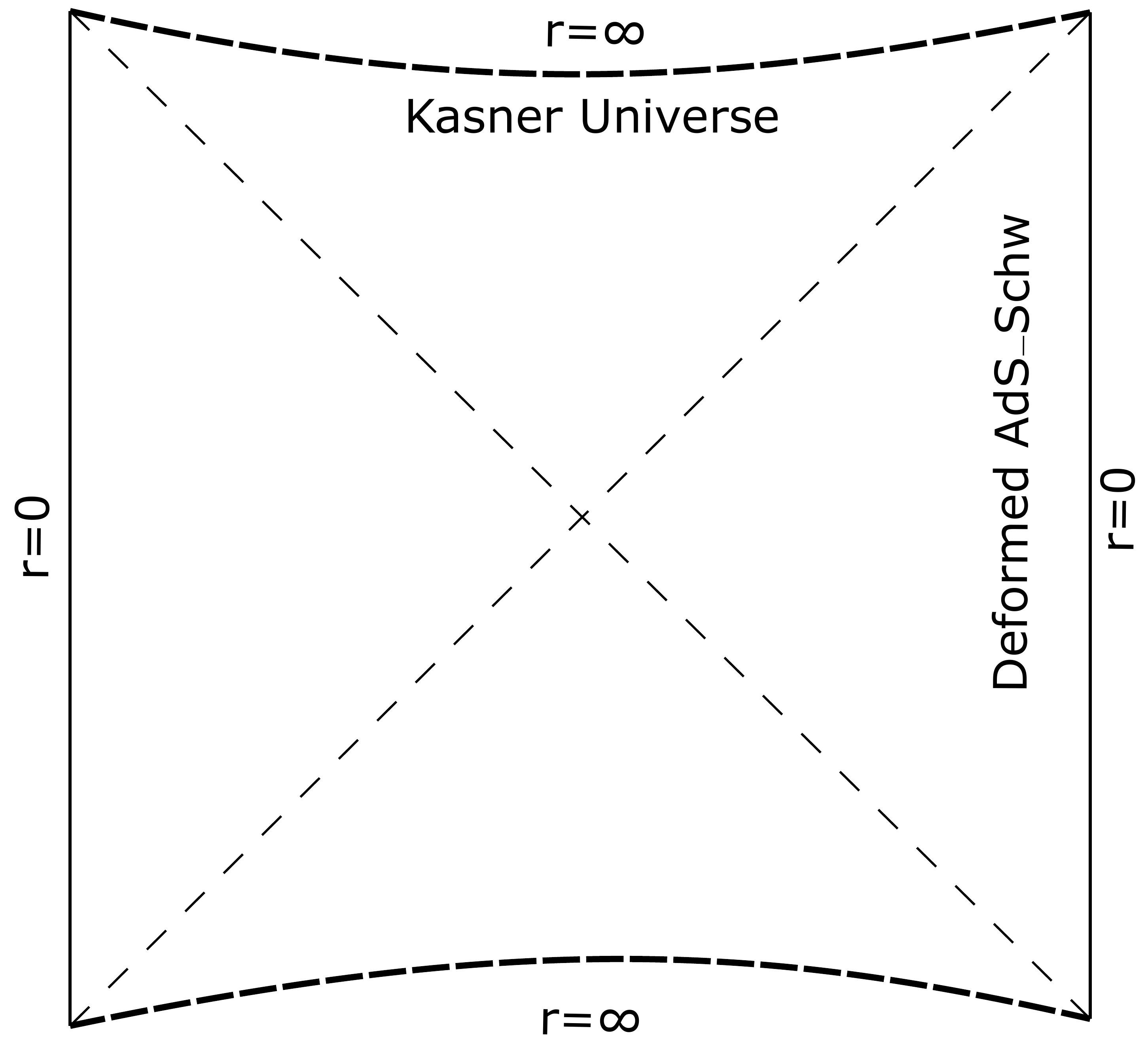}
    \caption{Deformed AdS-Schwarzschild solution at finite temperature $T$. The scalar field drastically alters the interior geometry from AdS-Schwarzschild to a more general Kasner universe near the spacelike singularity.}
    \label{fig:enter-label}
\end{figure}

The bulk equations of motion (EOM) are the usual Einstein's equations coupled to the scalar field stress-energy tensor, and the Klein-Gordon equation,
\begin{align}
G_{\mu\nu} - \frac{d(d-1)}{2}g_{\mu\nu} &= \frac{1}{4}\left[2\nabla_\mu \phi \nabla_\nu \phi - g_{\mu\nu}\left(\nabla^\alpha \phi \nabla_\alpha \phi + m^2\phi^2\right)\right]\,,\label{EinsteinM}\\
(\square - m^2)\phi &= 0\,,
\label{EinsteinS}
\end{align}
where $G_{\mu\nu}$ is the Einstein tensor.
We choose a particular ansatz for the metric,
\begin{equation}
ds^2 = \frac{1}{r^2}\left[-f(r)e^{-\chi(r)} dt^2 + \frac{dr^2}{f(r)} + d\vec{x}^2\right]\,,\label{metAnsatz}
\end{equation}
where $t \in \mathbb{R}$, $r > 0$, and $\vec{x} \in \mathbb{R}^{d-1}$. The conformal boundary is located at $r=0$ while the singularity is at $r=\infty$. The function $f(r)$ is known as the blackening factor, having a simple root at the black hole horizon $r=r_h$. Furthermore, we consider a profile for the scalar field depending only on the radial coordinate, $\phi = \phi(r)$. The dual scalar operator $\mathcal{O}$ corresponds to a homogeneous boundary deformation which, according to the AdS/CFT dictionary, has conformal dimension $\Delta$ satisfying
\begin{equation}
m^2 = \Delta(\Delta - d)\,.\label{massdimension}
\end{equation}
Plugging the ansatz into the EOM we obtain,
\begin{align}
\phi'' + \left(\frac{f'}{f} - \frac{d-1}{r} - \frac{\chi'}{2}\right)\phi' + \frac{\Delta(d-\Delta)}{r^2 f}\phi& = 0\,,\label{eom1}\\
\chi' - \frac{2f'}{f} - \frac{\Delta(d-\Delta)\phi^2}{(d-1)rf} - \frac{2d}{rf} + \frac{2d}{r} &= 0\,,\label{eom2}\\
\chi' - \frac{r}{d-1}\left(\phi'\right)^2 &= 0\,.\label{eom3}
\end{align}
The metric in \eqref{metAnsatz} must be continuous and regular at the horizon. To emphasize this point, it is convenient to switch to infalling Eddington-Finkelstein coordinates in which the metric takes the form,
\begin{equation}
ds^2 = \frac{1}{r^2}\left[-f(r) e^{-\chi(r)} du^2 + 2e^{-\chi(r)/2} du\,dr + d\vec{x}^2\right]\,.
\end{equation}
Note that AdS-Schwarzschild is the vacuum solution with no backreaction from $\phi$. This solution corresponds to $f(r) = 1-(r/r_h)^{d}$, $\chi(r) = 0$ and $\phi(r)=0$. The scalar field deforms slightly the exterior geometry but induces a large backreaction in the interior geometry, where $\phi(r)$ grows without bound. In particular, it leads to geometry that takes the form of a more general Kasner universe near the spacelike singularity. See Fig.~\ref{fig:enter-label} for an illustration. 

We now describe the behavior of the radial profiles and the corresponding asymptotic data at near the UV boundary ($r \to 0$) and the IR singularity ($r \to \infty$). Moreover, we discuss how the bulk represents an RG flow which interpolates from one to the other, allowing us to treat the near-singularity data as \textit{emergent} from the near boundary data. We briefly explain how to obtain these numerically.

The solutions are obtained by a shooting method, where the radial functions are integrated from the horizon $r = r_h$ to both the boundary $r \to 0$ and the singularity $r \to \infty$. By assuming regularity at the horizon, we can expand $\phi$, $f$, and $\chi$ as follows,
\begin{align}
\phi(r) &= \phi_h + \phi'_h (r-r_h) + O[(r-r_h)^2]\,,\\
f(r) &= f'_h (r-r_h) + O[(r-r_h)^2]\,,\\
\chi(r) &= \chi_h + \chi'_h (r-r_h) + O[(r-r_h)^2]\,.
\end{align}
The subscript $h$ denotes the values of the fields at the horizon. Plugging these expansions into the equations of motion \eqref{eom1}-\eqref{eom3} and taking the limit $r \to r_h$, we find the following constraints on the series coefficients,
\begin{align}
0 &= \frac{\Delta(d-\Delta)\phi_h}{r_h} + r_h f'_h \phi'_h\,,\\
0 &= -\frac{\Delta(d-\Delta)\phi_h^2}{d-1} - 2(d + r_h f'_h)\,,\\
0 &= \frac{r_h (\phi'_h)^2}{d-1} - \chi'_h\,.
\end{align}
Solving these equations we obtain,
\begin{align}
\phi_h &= \mp \frac{i\sqrt{2}\sqrt{d-1}\sqrt{d+f'_h r_h}}{\sqrt{\Delta(d-\Delta)}}\,,\\
\phi'_h &= \pm\frac{i\sqrt{2}\sqrt{d-1}\sqrt{d+f'_h r_h}\sqrt{\Delta(d-\Delta)}}{f'_h r_h^2}\,,\\
\chi'_h &= -\frac{2(d+f'_h r_h)\left[\Delta(d-\Delta)\right]}{f'^2_h r_h^3}\,.
\end{align}
Despite having a solution for these coefficients, we have the freedom to set a scale by fixing $f'_h$ numerically as long as it remains negative. This scale will later set the temperature of the black hole. We can further set $\chi_h = 0$. This is a gauge choice that 
fixes a normalization of the time coordinate and it is allowed because a shift in the $\chi$ field does not affect the equations of motion. Then, for each value of $r_h$ and taking some comparatively small $\delta > 0$, we can integrate the radial functions either from $r = r_h - \delta$ (outside of the horizon) to the boundary \textit{or} from $r = r_h + \delta$ (inside of the horizon) to the singularity.
To solve these equations numerically, we need to specify boundary conditions. Let us first explain the near-boundary data, and then we will explain the near-singularity data.
We can generically consider the near-boundary ($r \to 0$) mode expansion of a scalar field in AdS$_{d+1}$ to be,
\begin{equation}
\phi(r) \sim \phi_- r^{\Delta_-} + \phi_+ r^{\Delta_+}\,,\label{modeExpScal}
\end{equation}
where $\Delta_{\pm}$ are the two roots of \eqref{massdimension},
\begin{equation}
\Delta_{\pm} = \frac{1}{2}\left(d \pm \sqrt{d^2 + 4m^2}\right)\,.\label{roots}
\end{equation}
Further, from the mass-dimension relation \eqref{massdimension}, one can check that the dual operator will be relevant 
\begin{equation}
\Delta < d\,,
\end{equation}
if and only if $m^2 < 0$. However, we must also consider the Breitenlohner-Freedman stability bound \cite{Breitenlohner:1982bm,Breitenlohner:1982jf}, which translates into a lower bound for $m^2$,
\begin{equation}
-\frac{d^2}{4} \leq m^2 < 0\,.
\end{equation}
Meanwhile the roots \eqref{roots} satisfy,
\begin{equation}
0 < \Delta_- \leq \frac{d}{2} \leq \Delta_+ < d\,,
\end{equation}
with $\Delta_\pm = d/2$ when $m^2 = -d^2/4$.\footnote{Note that when $m^2 = -d^2/4$ the expansion \eqref{modeExpScal} degenerates and must be generalized.} 
Based on the value of $m^2$, there may be a choice to be made for which root is taken to be $\Delta$ (related to the boundary conditions of $\phi$). For $m^2 \geq 1-d^2/4$, the only option is $\Delta = \Delta_+$ because selecting $\Delta = \Delta_-$ would violate the unitarity bound, $\Delta > (d-2)/2$.
However, $m^2 < 1 - d^2/4$ gives us both options. While the canonical choice $\Delta = \Delta_+$ restricts us to $\Delta \geq d/2$ (stricter than unitarity), using the alternative quantization for which $\Delta = \Delta_-$ allows us to reach as low as $(d-2)/2$. So for a bulk theory with a scalar field with $-d^2/4 < m^2 < 1-d^2/4$, we have \textit{two} possible dual CFTs ---one with $\Delta = \Delta_+$ and one with $\Delta = \Delta_-$--- related by a Legendre transform of the generating functional.

Working in the standard quantization, where $\Delta = \Delta_+$, we can now express the leading-order and next-to-leading-order coefficients in \eqref{modeExpScal} in terms of the source and expectation value of the dual operator, $\phi_0$ and $\braket{\mathcal{O}}$, respectively,
\begin{equation}
\phi(r) \sim \phi_0 r^{\Delta_-} + \frac{\braket{\mathcal{O}}}{2\Delta_+ - d}r^{\Delta_+}\,.
\end{equation}
However, if we choose to work in the alternative quantization, this identification is flipped. In this case, $\phi_+$ is identified as the source and $\phi_-$ as the expectation value. Either way, a generic way to write the near-boundary expansion is as follows,
\begin{equation}
\phi(r) \sim \phi_0 r^{d-\Delta} + \frac{\braket{\mathcal{O}}}{2\Delta - d}r^{\Delta}\,,
\end{equation}
since $\Delta_+ + \Delta_- = d$. As for the particular case $\Delta = d/2$, a logarithmic divergence appears in the $r^\Delta$ term, making the breakdown of modes evident \cite{Minces:1999eg}. In particular, near the asymptotic boundary, the term proportional to the source behaves as
\begin{equation}
\phi(r) \sim \phi_0 r^{d/2}\log r\,.\label{logPhi}
\end{equation}

By integrating the EOM from the horizon to the boundary, we obtain $\phi(r)$ in the exterior. The scalar field profile is used to get $\phi_0$, but how we do so depends on whether we are working in the standard or alternative quantization. This is because the power of the source term is only leading in the former case. In short, we find,
\begin{equation}
\phi_0 = \begin{cases}
\displaystyle\lim_{r \to 0} r^{\Delta - d}\phi(r)\,,& \Delta > \dfrac{d}{2}\,,\\\\[-3ex]
\displaystyle\lim_{r \to 0} -\dfrac{r^{2\Delta - d + 1}}{2\Delta - d}\partial_r \left[r^{-\Delta}\phi(r)\right]\,,& \Delta < \dfrac{d}{2}\,.
\end{cases}
\end{equation}
As expected, neither of these formulas gives the correct result for $\Delta = d/2$. In this case we use the logarithmic expression (\ref{logPhi}) to obtain
\begin{equation}
\phi_0 = \lim_{r \to 0} \frac{r^{-d/2}}{\log r} \phi(r),\ \ \Delta = \frac{d}{2}\,.
\end{equation}

Similarly, we can obtain $\chi(r)$ in the exterior. Ideally, we would like to obtain a solution where $\chi(0)=0$ so that the boundary time is canonically normalized. However, since we have set $\chi_h = 0$ in order to do the integration, this boundary value will not be guaranteed in general. Since a constant shift in $\chi$ does not affect the equations of motion, there is a simple fix to this problem. By simply evaluating $\chi(0)$ and shifting the entire function by this amount, we can obtain the `true' $\chi(r)$ for which $\chi(0) = 0$. In doing so, we also get the `true' $\chi_h$, which, in combination with $f'_h$ fixes the temperature of the black hole,
\begin{equation}
    T=\frac{1}{\beta}=\frac{|f'_h|e^{-{\chi_h}/2}}{4\pi}\,.
\end{equation}

\begin{figure}
		\centering
		\begin{subfigure}{0.48\textwidth}
			\centering
			\includegraphics[width=\textwidth]{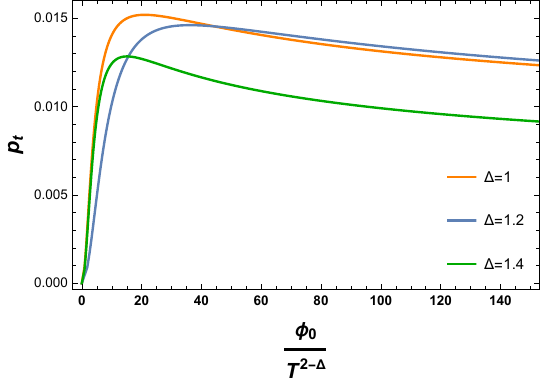}
		\end{subfigure}
		\hfill
		\begin{subfigure}{0.478\textwidth}
			\centering
			\includegraphics[width=\textwidth]{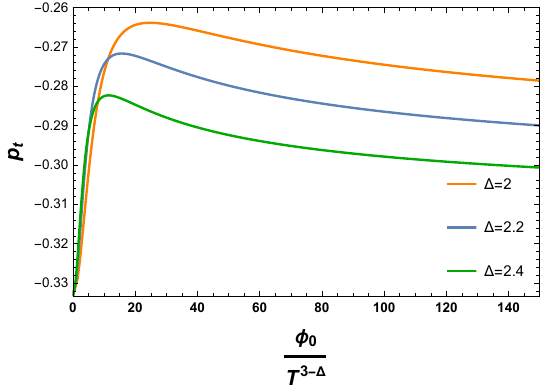}
		\end{subfigure}
		\caption{The Kasner exponent $p_t$ as a function of boundary deformation $\frac{\phi_0}{T^{d-\Delta}}$ for $d=2$ (left) and $d=3$ (right), and various values of $\Delta$.}
		\label{ptvsphi}
\end{figure}

In the interior, the fields generically diverge as they approach the singularity,
\begin{equation}
    \phi(r)\sim (d-1)c\log r\,,\,\,\,\,\,\,\,\,\chi(r)\sim (d-1)c^2 \log r+\chi_1\,,\,\,\,\,\,\,\,\,\, f(r)\sim-f_1 r^{\big(d+\frac{d-1}{2}c^2\big)}\,,
\end{equation}
where $c$, $\chi_1$ and $f_1$ are the near-singularity constants. After plugging these solutions into (\ref{metAnsatz}) and redefining the radial coordinate as $\tau=r^{-\frac{1}{2}\big({d+\frac{d-1}{2}c^2}\big)}$, we find an isotropic Kasner universe near the singularity,
\begin{equation}
    ds^2\sim -d\tau^2+\tau^{2 p_t}dt^2+\tau^{2 p_x}d\vec{x}^2\,,\,\,\,\,\,\,\,\phi(\tau)\sim -\sqrt{2}p_\phi \log{\tau}\,.
\end{equation}
The Kasner exponents are given by,
\begin{equation}
    p_t=\frac{(d-1)c^2-2(d-2)}{2d+(d-1)c^2}\,,\quad p_x=\frac{4}{2d+(d-1)c^2}\,,\quad p_\phi=\frac{2\sqrt{2}(d-1)c}{2d+(d-1)c^2}\,,\label{eq:kasnerc}
\end{equation}
satisfying the constraints,
\begin{eqnarray}
    p_t+(d-1)p_x&=&1\,,\\
    p_\phi^2+p_t^2+(d-1)p_x^2&=&1\,.
\end{eqnarray}
The constraints imply that only one of these exponents is independent. When integrating to the singularity, we obtain $\phi(r)$, from which we can get the constant $c$ and hence the Kasner exponents. In this way, we can generate a plot between the deformation parameter (normalized with respect to the temperature), $\frac{\phi_0}{T^{d-\Delta}}$, and the desired Kasner exponent. For example, the end result for $p_t$ vs. $\frac{\phi_0}{T^{d-\Delta}}$ is shown in Fig.~\ref{ptvsphi}.

\section{Perturbing RG flows with shock waves}{\label{Rg with a shock}}
We will now perturb the deformed geometries with a shock wave. The shock wave carries energy $E$ which we assume to be smaller than the mass of the black hole $M$. Nevertheless, it increases the mass of the black hole by some amount from $M$ to $M+E$. In order to understand its potential impact on the black hole's internal dynamics, we will explore several observables: out-of-time-order correlators computed in the
heavy-heavy-light-light limit, the thermal $a$-function, and the entanglement velocity.

\subsection{Geodesic length and OTOCs\label{sec:geo}}
One of the most famous entries of the AdS/CFT dictionary relates two-point correlators in the boundary theory to certain bulk paths connecting the two insertion points on the boundary theory. More concretely, in `first quantized' language, the dictionary stipulates that for operators $\mathcal{O}$ with conformal dimension $\Delta_{\mathcal{O}}$ \cite{Balasubramanian:1999zv,Louko:2000tp},
\begin{equation}
   \braket{{\mathcal{O}}(X){\mathcal{O}}(Y)}=\int_{X}^{Y} d \mathcal{P} e^{-\Delta_{\mathcal{O}} L(\mathcal{P})}\,,\label{eq:geoapprox}
\end{equation}
where $\mathcal{P}$ is the all possible paths connecting the two points $X$ and $Y$ and $L(\mathcal{P})$ represents the lengths of those paths. In the large mass, or large conformal dimension limit, $\Delta_{\mathcal{O}} \to \infty$, the correlation function can be well approximated by the sum of the geodesic lengths connecting the two points,
\begin{equation}
   \braket{{\mathcal{O}}(X){\mathcal{O}}(Y)}\approx\sum_{\text{geodesics}}e^{-\Delta_{\mathcal{O}} L(X,Y)} \,,
   \label{twopointc}
\end{equation}
where $L(X,Y)$ is now geodesic length between $X$ and $Y$.

We are interested in computing the 2-point correlation functions between symmetrically placed boundary points on the shockwave geometries, so, the corresponding bulk objects of interest are spacelike geodesics anchored at those boundary endpoints.\footnote{In certain cases, complex saddles may dominate the path integral (\ref{eq:geoapprox}) \cite{Fidkowski:2003nf}. Particularly, in the presence of a shock wave, complex geodesics turn out to be crucial for computing the late time position space correlators \cite{Horowitz:2023ury}. However, real geodesics do play a role in computing momentum space/hybrid correlators. We elaborate on this subtle point in section \ref{sec:hybrid}.} As it is well known, those correlation functions can be related by analytic continuation to the four-point out-of-time order correlators (OTOCs) that diagnose quantum chaos, in the so-called heavy-heavy-light-light limit \cite{Shenker:2013pqa}. See \cite{Jahnke:2018off} for a review. With this computation, then, we will be able to explore chaotic properties of the holographic RG flows and explore possible imprints from and on the Kasner interiors.

\begin{figure}
		\centering
		\begin{subfigure}{0.45\textwidth}
			\centering
			\includegraphics[width=\textwidth]{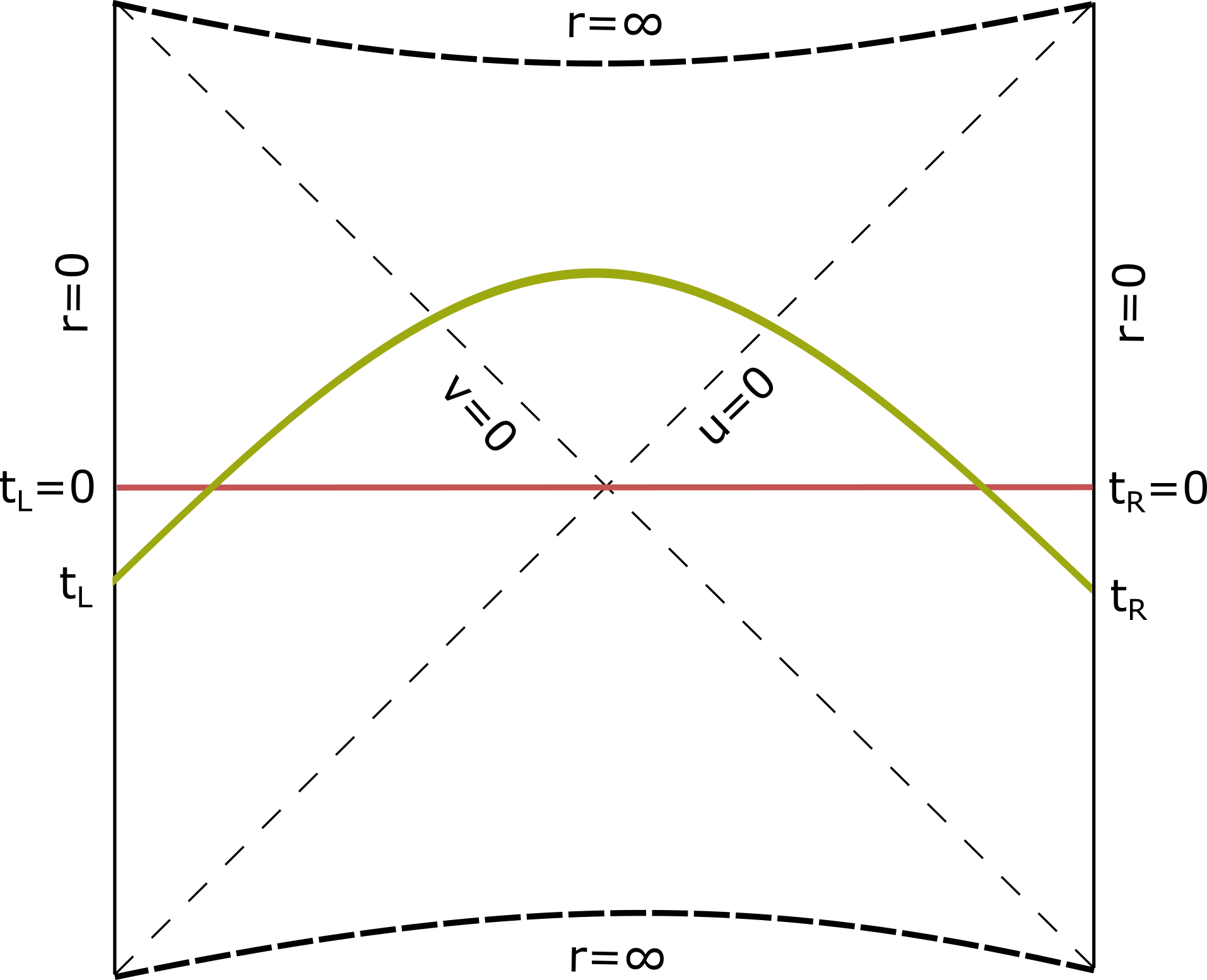}
		\end{subfigure}
		\hfill
		\begin{subfigure}{0.49\textwidth}
			\centering
			\includegraphics[width=\textwidth]{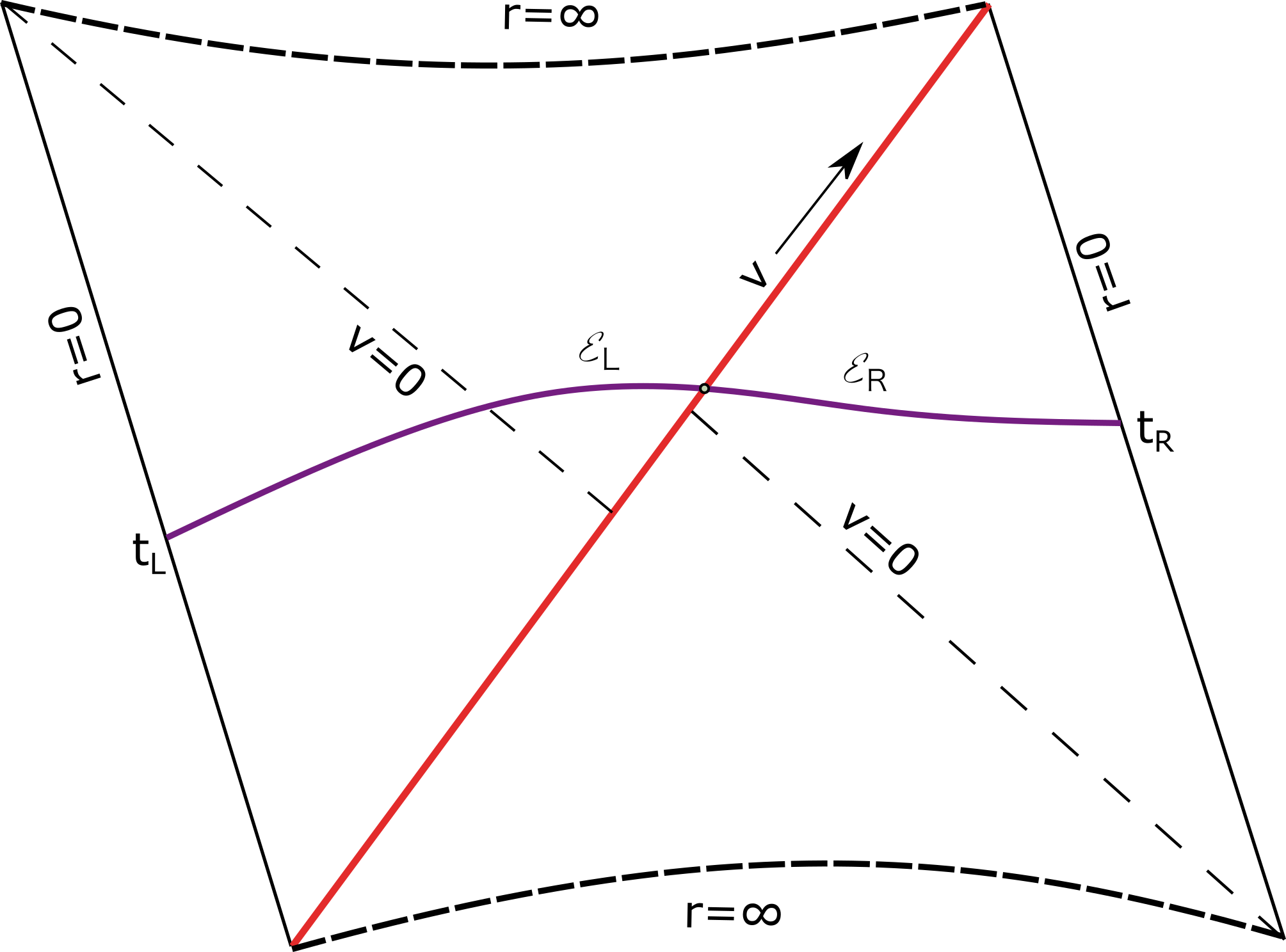}
		\end{subfigure}
		\caption{Left: spacelike geodesics on a generic black hole geometry (other than BTZ). The solid red line represents the minimal geodesic at $t_L=t_R=0$. At some other boundary time $t_L=t_R=t_b$, the minimal geodesic is instead given by the solid green line. If $t_b>0$ the geodesics bend towards the past. Conversely, for $t_b<0$ the geodesics bend towards the future \cite{Fidkowski:2003nf}. Right: spacelike geodesics on the shock wave geometry. The shock is sent at a very early time from the left boundary, shifting the $v$ coordinate by a constant amount $\alpha$ along the shock. The solid purple line in this geometry represents a geodesic anchored at some boundary time $t_L=t_R=t_b$.}
		\label{bhwithshock}
\end{figure}

In a previous work \cite{Frenkel:2020ysx}, the authors computed these spacelike geodesics for a metric of the form (\ref{metAnsatz}), without a shock wave perturbation. See Fig.~\ref{bhwithshock} (left) for an illustration. Since the shock wave geometry is still given in the form (\ref{metAnsatz}), but piecewise,\footnote{More concretely, one side of the shock wave is a black hole of the form (\ref{metAnsatz}) with mass $M$ and the other side a black hole of the same form with mass $M+E$.} it will be useful to review their calculation. 

To start, it is important to note that due to the spatial symmetry, these geodesics will lie on a constant $\vec{x}$ plane. The induced metric on such a plane is,
\begin{equation}
ds^2|_{\vec{x}=\text{constant}}=\frac{1}{r^2}\left[-f(r)e^{-\chi(r)} dt^2 + \frac{dr^2}{f(r)}\right]\,.
\end{equation}
The spacelike geodesics on this plane can be parametrized as $r=r(t)$ so that their length functional becomes,
\begin{equation}
    L=\int \frac{dt}{r}\left[-f(r)e^{-\chi(r)}+\frac{\dot{r}^2}{f(r)}\right]^{{1}/{2}}=\int dt \mathcal{L}\label{geol}\,,
\end{equation}
where $\mathcal{L}$ is the Lagrangian,
\begin{equation}
    \mathcal{L}=\frac{1}{r}\sqrt{-f(r)e^{-\chi(r)}+\frac{\dot{r}^2}{f(r)}}\,.
\end{equation}
Note that since $\mathcal{L}$ does not depend explicitly on time, we can define a conserved quantity $\mathcal{E}$, which represents the energy associated with the spacelike geodesic,
\begin{equation}
    \mathcal{E}=\dot{r}\frac{\partial\mathcal{L}}{\partial\dot{r}}-\mathcal{L}=\frac{f(r)e^{-\chi(r)}}{r\sqrt{-f(r)e^{-\chi(r)}+\frac{\dot{r}^2}{f(r)}}}\,.
    \label{constom}
\end{equation}
The geodesic length can then be obtained by plugging (\ref{constom}) into (\ref{geol}). This yields the minimal geodesic length anchored at some boundary time slice $t=t_b(\mathcal{E})$,
\begin{equation}
    L=\frac{2}{|\mathcal{E}|}\int_{r_c}^{r_t} \frac{e^{-\chi/2}dr}{r^2\sqrt{1+\frac{f(r)e^{-\chi(r)}}{r^2\mathcal{E}^2}}}+2 \log r_c\,.\label{emptylength}
\end{equation}
Note we have added a counterterm to subtract the UV divergences coming from the AdS boundary. Here $r_c$ represents the UV cutoff and  $r_t$ is the turning point of the geodesic, determined by the relation, 
\begin{equation}
\mathcal{E}^2=-\frac{f(r_t)e^{-\chi(r_t)}}{r_t^2}\,.
\end{equation}
The boundary time can likewise be expressed as a function of the turning point $r_t$ (or, equivalently, as a function of $\mathcal{E}$),
\begin{equation}
    t_b=-\int_{0}^{r_t}\frac{dr}{\dot{r}}=-\int_{0}^{r_t}\frac{e^{\chi/2}dr}{f(r)\sqrt{1+{f(r)e^{-\chi(r)}}/(r \mathcal{E})^2}}\,.
    \label{boundarytime}
\end{equation}
Finally, one can relate the geodesic length to the boundary time by expressing both as a function of the turning point $r_t$ and studying their parametric dependences.

Let us now move on to the shock wave geometry, which we depict in Fig.~\ref{bhwithshock} (right).
The shock wave is sent from the left asymptotic boundary at some late boundary time $t_w$. The energy of the shock wave, which we denote as $E$, is significantly small compared to the mass of the black hole, $M$. However, it is enough to increase the mass of the black hole from $M$ to $M+E$. Our focus is to compute the two-point function in the boundary theory, which for the case of heavy operators, is given by the minimal geodesic length on this perturbed geometry. To achieve this, we split the calculation into two steps. First, we compute the length of all possible geodesics anchored at one of the boundaries that intersect the shock at a given point. Second, we extremize over all possible intersections along the shock wave.

In order to proceed, we perform a coordinate transformation from the standard Schwarzschild coordinates (\ref{metAnsatz}) to the fully extended Kruskal coordinates $(u,v)$,
\begin{equation}\label{Kcoords}
    uv=-e^{\frac{4\pi}{\beta}r_*(r)}\,,\quad u/v=-e^{-\frac{4\pi}{\beta}t}\,,
\end{equation}
where, 
\begin{equation}
    t(r)=t_b+\mathcal{E}\int\frac{r e^{\chi/2} dr}{f(r)\sqrt{r^2 \mathcal{E}^2+f(r)e^{-\chi}}}\,,\quad r_*(r)=\int\frac{e^{\chi/2}dr}{f(r)}\,.
\end{equation}
In the Penrose diagram, the shock wave propagates along a null surface at constant $u$. In the limit $E\ll M$, the net effect of the shock amounts to a shift in the $v$ coordinate by a constant amount $\alpha$, without affecting the other coordinate $u$. To make this more precise, it is convenient to use two different sets of Kruskal coordinates, $(u,v)$ for the right (or past) of the shock, and $(\tilde{u},\tilde{v})$ for the left (or future) of the shock.  The null shell propagates along the surface,
\begin{equation}
    \Tilde{u}_w=e^{\frac{2\pi}{\Tilde{\beta}}(\Tilde{r}_*(0)-t_w)}\,,\quad {u}_w=e^{\frac{2\pi}{{\beta}}({r}_*(0)-t_w)}\,.\label{shocktraject}
\end{equation}
Meanwhile, by ensuring that the metric across the shock
wave stays continuous imposes the following matching conditions:
\begin{equation}
    \Tilde{u}_w \Tilde{v}=-e^{\frac{4\pi}{\Tilde{\beta}}\Tilde{r}_*(r)}\,,\quad u_w v=-e^{\frac{4\pi}{\beta}r_*(r)}\,.\label{shockmatching}
\end{equation}
For small enough shock wave energy $E$, one can approximate $\tilde{u}_w=u_w$. At the same time, for large values of $t_w$ the value of radial direction $r$ is pushed towards the horizon radius $r_h$. Thus we can expand the fields around the horizon, $\xi(r)=f(r)e^{-\chi(r)/2}=\xi'(r)(r-r_h)+...=\frac{4\pi}{\beta}(r-r_h)$. By evaluating $r_*$, we further find $e^{\frac{4\pi}{\beta}r_*(r)}=\mathcal{C}(r,r_h)(r-r_h)$, where $\mathcal{C}$ is a smooth function and non-zero at the horizon. Then we find that the shift $\alpha$ in the small $E$ and large $t_w$ limit is given by,\footnote{See Appendix B of \cite{Shenker:2013pqa}.}
\begin{equation}
    \alpha=\tilde{v}-v=\frac{E}{u_w}\frac{d r_h}{dM}\mathcal{C}(r_h,r_h)\,.
    \label{shockshift}
\end{equation}

We now proceed to split the geodesic into two pieces $L=L_1+L_2$ and compute their corresponding lengths. Namely, we compute the geodesic length from the right boundary up to the shock and then from the shock to the left boundary. At the end of the calculation, we extremize the sum with respect to the intersection point along the shock wave. When the energy of the right geodesic has $\mathcal{E}_R>0$, we solve the differential equation for $v_R$ to determine its value as a function of the conserved quantity $\mathcal{E}_R$.
On the other hand, for $\mathcal{E}_R<0$, we solve for $u_R$ instead of $v_R$. Note that in this case the geodesic has a turning point before it intersects the shock wave. In summary, we have\footnote{For $\mathcal{E}_R<0$ we must multiply $u$ by $\exp({-2r_*})$ to determine $v$ at the horizon.}
\begin{eqnarray}
    \frac{v'(r)}{v}-\frac{2\pi e^{\chi/2}}{\beta}\frac{\bigg(r\mathcal{E}_R-\sqrt{f(r)e^{-\chi/2}+r^2\mathcal{E}_R^2}\bigg)}{f(r)\sqrt{f(r)e^{-\chi/2}+r^2\mathcal{E}_R^2}}&=&0\,,\,\,\,\,\,\,\,\,\,\,\,\,\,\,\,\, \mathcal{E}_R>0\,,\label{rightvr1}\\
  \frac{u'(r)}{u}-\frac{2\pi e^{\chi/2}}{\beta}\frac{\bigg(r\mathcal{E}_R+\sqrt{f(r)e^{-\chi/2}+r^2\mathcal{E}_R^2}\bigg)}{f(r)\sqrt{f(r)e^{-\chi/2}+r^2\mathcal{E}_R^2}}&=&0\,,\,\,\,\,\,\,\,\,\,\,\,\,\,\,\,\, \mathcal{E}_R<0\,.
    \label{rightvr2}
\end{eqnarray}
Similarly, for the left geodesic we must solve
\begin{eqnarray}
    \frac{\tilde{u}'(r)}{\tilde{u}}+\frac{2\pi e^{\chi/2}}{\beta}\frac{\bigg(r\mathcal{E}_L-\sqrt{f(r)e^{-\chi/2}+r^2\mathcal{E}_L^2}\bigg)}{f(r)\sqrt{f(r)e^{-\chi/2}+r^2\mathcal{E}_L^2}}&=&0\,,\,\,\,\,\,\,\,\,\,\,\,\,\,\,\,\, \mathcal{E}_L<0\,,\\
    \frac{\tilde{v}'(r)}{\tilde{v}}+\frac{2\pi e^{\chi/2}}{\beta}\frac{\bigg(r\mathcal{E}_L+\sqrt{f(r)e^{-\chi/2}+r^2\mathcal{E}_L^2}\bigg)}{f(r)\sqrt{f(r)e^{-\chi/2}+r^2\mathcal{E}_L^2}}&=&0\,\,\,\,\,\,\,\,\,\,\,\,\,\,\,\,\, \mathcal{E}_L>0\,.
    \label{leftvl}
\end{eqnarray}
For a better illustration of the procedure, we show a geodesic in this backreacted geometry in Fig.~\ref{bhwithshock} (right). 

The total length $L$ is given sum of the two geodesic lengths. Assuming $\mathcal{E}_R>0$ and $\mathcal{E}_L<0$, so that the turning point is on the left (future) of the shock wave, we find
\begin{eqnarray}
L_1&=&\frac{1}{|\mathcal{E}_L|}\int_{r_c}^{r_t} \frac{e^{-\chi/2}dr}{r^2\sqrt{1+\frac{f(r)e^{-\chi}}{r^2\mathcal{E}_L^2}}}+\frac{1}{|\mathcal{E}_L|}\int_{r_h}^{r_t}\frac{e^{-\chi/2}dr}{r^2\sqrt{1+\frac{f(r)e^{-\chi}}{r^2\mathcal{E}_L^2}}}\,,\\
L_2&=& \frac{1}{|\mathcal{E}_R|}\int_{r_c}^{r_h}\frac{e^{-\chi/2}dr}{r^2\sqrt{1+\frac{f(r)e^{-\chi}}{r^2\mathcal{E}_R^2}}}\,,\\
L&=&L_1+L_2\,.
\end{eqnarray}
The left/right energies $\mathcal{E}_L$ and $\mathcal{E}_R$ can be expressed as a function of $v$ and $\tilde{v}$ at the horizon. Moreover, the shock wave shifts the $v$ coordinate by a constant amount, i.e. $\tilde{v}=v+\alpha$. Thus, we can establish a relation between $\mathcal{E}_L$ and $\mathcal{E}_R$ by using this shift in $v$. This means that the total length $L$ can be written solely as a function of $\mathcal{E}_R$. Finally, by extremizing with respect to this energy, we can ultimately derive the geodesic length. 

\subsubsection*{Example of the procedure: shock waves in BTZ}

A simple example in which we can carry out the above steps explicitly is that of the BTZ black hole. This solution is characterized by $f(r)=(1-r^2/r_h^2)$, $\chi(r)=0$ and $\phi(r)=0$.
By solving (\ref{rightvr1})-(\ref{leftvl}) in this background, we can get the following relation between the energies $\mathcal{E}_{R,L}$ and the values of $v$ and $\tilde{v}$ at the horizon,
\begin{eqnarray}
   \mathcal{E}_R&=&\frac{e^{-t_R/r_h}v}{r_h(1-e^{-t_R/r_h}v)}\,,\\
   \mathcal{E}_L&=-&\frac{e^{-t_L/r_h}\tilde{v}}{r_h(1-e^{-t_L/r_h}\tilde{v})}\,,
\end{eqnarray}
where $t_R$ and $t_L$ are the right and left anchoring times of the geodesic. Plugging these values for the energies in the total length and then extremizing with respect to $\mathcal{E}_R$, we find the total (regulated) length of the extremal geodesic,
\begin{equation}
    L=L_1+L_2=2\log{{2r_h}+2\log\left(\cosh\bigg(\frac{t_L+t_R}{2 r_h}\bigg)+\frac{\alpha}{2}e^{-(t_R-t_L)/2r_h}\right)}\,.
\end{equation}
This matches with the expected result for the geodesic length, derived in \cite{Shenker:2013pqa}. 

\subsubsection*{Relation to quantum chaos}

In order to understand the chaotic effects of the shock wave on the system, it suffices to set $t_L=t_R=0$ and study the response of the two-point function (\ref{twopointc}) to a perturbation sent at a very early time $t_w$. For example, in the case of a BTZ black hole, the shock wave parameter $\alpha$ (\ref{shockshift}) simplifies to
\begin{equation}
    \alpha=\frac{E}{4 M} e^{t_w/r_h}\,.
\end{equation}
From (\ref{twopointc}), it follows that the (normalized) two-point function on the shock wave geometry is,
\begin{eqnarray}
   \frac{\braket{{\mathcal{O}}(t_L){\mathcal{O}}(t_R)}_w}{\braket{{\mathcal{O}}(t_L){\mathcal{O}}(t_R)}_{0}}\bigg|_{t_L=t_R=0} &=& \frac{e^{-\Delta_{\mathcal{O}} L_w(0,0)}}{e^{-\Delta_{\mathcal{O}} L_0(0,0)}} =\bigg(\frac{1}{1+\frac{E}{8 M}e^{t_w/r_h}}\bigg)^{2\Delta_{\mathcal{O}}}\label{QCBeffect0}\\
   &\simeq&1-\frac{\Delta_{\mathcal{O}} E}{4M}e^{\frac{t_w}{r_h}}+O(E^2/M^2) \,,\label{QCBeffect}
\end{eqnarray}
where $L_w$ and $L_0$ represent the geodesic lengths in the presence and absence of the shock wave, $r_h=\frac{\beta}{2\pi}$ is the black hole horizon radius and $\beta$ is the inverse temperature. The latter approximation is valid when $t_\text{th}\ll t \ll t_*$, where $t_\text{th}$ is the scale of local thermalization $t_\text{th}\sim \beta$ and $t_*$ is the so-called scrambling time $t_*\sim\beta\log{S_{\text{BH}}}$. Typically, $t_\text{th}$ is set by the decay of local perturbations, which can be obtained from a quasi-normal mode analysis. Meanwhile, $t_*$ represents a more global scale, which measures the spread of information throughout $O(1)$ of the black hole's degrees of freedom \cite{Sekino:2008he}. 

The exponential decrease of correlations following local thermalization exemplifies the `butterfly effect,' often regarded as the smoking gun of quantum chaos.
In fact, upon analytic continuation, (\ref{QCBeffect0}) can be related to the infamous OTOC that diagnoses the exponential growth of the commutator squared $C(t)=-\langle [\mathcal{W}(t),\mathcal{O}(0)]^2\rangle$ for a generic pair of (Hermitian) operators, $\mathcal{W}$ and $\mathcal{O}$, inserted at the same boundary. More specifically, (\ref{QCBeffect0}) can be mapped to the OTOC
\begin{equation}
    F(t)=\frac{\braket{{\mathcal{O}}(0){\mathcal{W}}(t){\mathcal{O}(0)}{\mathcal{W}}(t)}}{\braket{{\mathcal{O}(0)}\mathcal{O}(0)}\braket{{\mathcal{W}}(t)\mathcal{W}(t)}}\,,\label{OTOCchaos}
\end{equation}
in the so-called heavy-heavy-light-light limit, where a pair of operators backreact on the geometry (the $\mathcal{W}$'s) and the other two probe the deformed geometry (the $\mathcal{O}$'s). For theories with a large number of degrees of freedom $N^2$, there is a parametrically large hierarchy between scrambling and thermalization times, and (\ref{OTOCchaos}) takes the form
\begin{equation}\label{generalOTOC}
    F(t)\sim 1-\frac{f_0}{N^2}e^{\lambda_L t}+\cdots\sim 1-e^{\lambda_L (t-t_*)}+\cdots\,,
\end{equation}
where $t_*=\frac{1}{\lambda_L}\log(\frac{N^2}{f_0})$ is the scrambling time and $\lambda_L$ is the (quantum) Lyapunov exponent. In general quantum systems, $\lambda_L$ is known to be bounded from above by \cite{Maldacena:2015waa}
\begin{equation}\label{cbound}
\lambda_L\leq\frac{2\pi}{\beta}\,.
\end{equation}
A particularly noteworthy result is that for holographic theories with Einstein gravity duals, this bound is precisely saturated, providing compelling evidence for the assertion that black holes stand as the fastest scramblers in nature \cite{Sekino:2008he}.

\subsubsection*{Chaotic OTOCs in holographic RG flows}

The gravitational sector of the RG flows studied in this paper is just pure Einstein gravity, so $\lambda_L$ must saturate the bound (\ref{cbound}). Moreover, the number of degrees of freedom in a holographic theory is set in terms of the AdS radius $L$, and Newton's constant $G$, $N^2\sim L^{d-1}/G$. Thus, at least in the time window of interest, $t_\text{th} \ll t \ll t_*$, all that is left to determine is the constant $f_0$ in (\ref{generalOTOC}). This will depend on the strength of the deformation $\phi_0$, or more specifically, on the dimensionless combination $\frac{\phi_0}{T^{d-\Delta}}$. 

\begin{figure}
    \centering
    \includegraphics{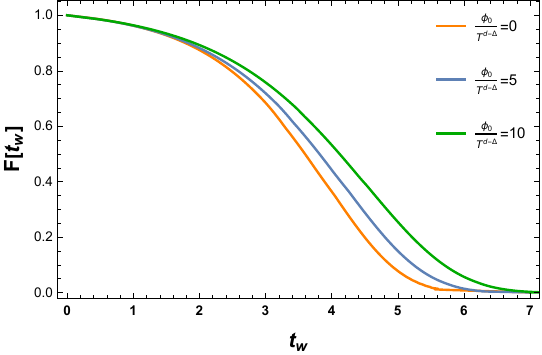}
    \caption{The OTOC as a function of shock time $t_w$ for $d=2$ and $\Delta=1.4$. We observe that the scrambling time $t_*$ increases as we increase the strength of the deformation, which we can estimate as the time at which $F(t)$ is decreased by an order one amount.}
    \label{perturbl}
\end{figure}

With the geodesic method outlined above, we can actually do a bit better. We can compute $F(t)$ all the way to the scrambling regime, where the OTOC is expected to fade away and eventually drop to zero. We show a sample of our results in Fig.~\ref{perturbl}. One can roughly estimate the scrambling time by observing when $F(t)$ is affected by order one amount. Based on Figure \ref{perturbl}, it can be inferred that the scrambling time increases as the strength of the relevant deformation $\frac{\phi_0}{T^{d-\Delta}}$ increases. We will provide a more quantitative analysis of this observable in the next subsection.

\subsubsection{Scrambling time}
To validate the claim of the previous section, here we will estimate the time that a shock sent at a very early time takes to scramble across the event horizon. More precisely,  we will determine $t_w$ ---the time at which the shock wave is inserted at the boundary--- such that it makes an $O(1)$ shift in $\alpha$ \cite{Shenker:2013pqa}. For that purpose, we will briefly recap some of the key steps in the construction of the shock wave geometries.

We start with a metric of the form (\ref{metAnsatz}),
and introduce two sets of Kruskal coordinates ($u,v$) and ($\tilde{u},\tilde{v}$), defined as in (\ref{Kcoords}). We then add a null perturbation with energy $E \ll M$ at a time $t_w$, sent from the left asymptotic boundary.  The coordinates ($u,v$) will describe the portion of the manifold to the right (past) of the shock while the ($\tilde{u},\tilde{v}$) coordinates ($u,v$) will describe the portion of the manifold to the right (future) of the shock. The null perturbation propagates along the surface (\ref{shocktraject}). Meanwhile, the matching conditions imply some relations between tilded and un-tilded variables (\ref{shockmatching}), from which one can read off a formula for the shift $\alpha$, given by (\ref{shockshift})
\begin{equation}
    \alpha=\frac{E}{u_w}\frac{d r_h}{dM}\mathcal{C}(r_h,r_h)\,.
\end{equation}
All quantities at the left are known, except for $\mathcal{C}(r_h,r_h)$. This is defined as the coefficient of the divergent term in $r_*$ as $r\to r_h$. More specifically, the function $\mathcal{C}(r,r_h)$ is defined as $\mathcal{C}(r,r_h)=e^{\frac{4\pi}{\beta}r_*(r)}/(r-r_h)$, which is finite since $r_*(r)$ diverges logarithmically near the horizon. Finally, by using the area law for the black hole horizon and the first law of thermodynamics, we finally determine the scrambling time after setting $\alpha=1$,
\begin{equation}
    t_*=r_*(0)+\frac{\beta}{2\pi}\log\left[\frac{(d-1)\text{Vol}(R^{d-1})T}{4 \mathcal{C}(r_h,r_h) E G_N r_h^{d}}\right]\,.\label{eq:scrambling}
\end{equation}

\begin{figure}
		\centering
		\begin{subfigure}{0.45\textwidth}
			\centering
			\hspace{-0.3cm}\includegraphics[trim={0 0 0 0},clip,width=\textwidth]{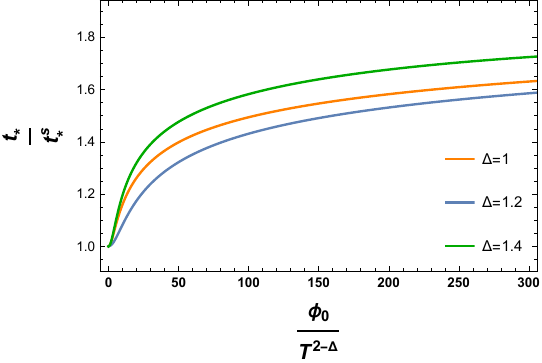}
		\end{subfigure}
		\hspace{0.2cm}
		\begin{subfigure}{0.45\textwidth}
			\centering
			\includegraphics[trim={0 -0.615cm 0 0},clip,width=\textwidth]{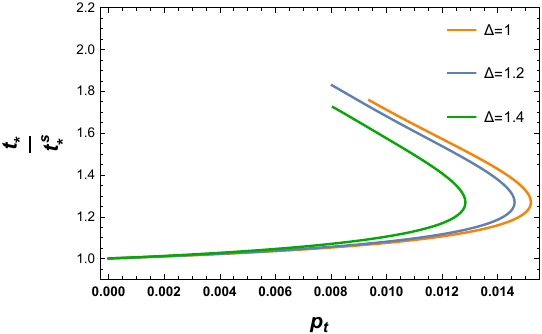}
		\end{subfigure}
  \begin{subfigure}{0.45\textwidth}
			\centering
			\hspace{-0.3cm}\includegraphics[trim={0 0 0 0},clip,width=\textwidth]{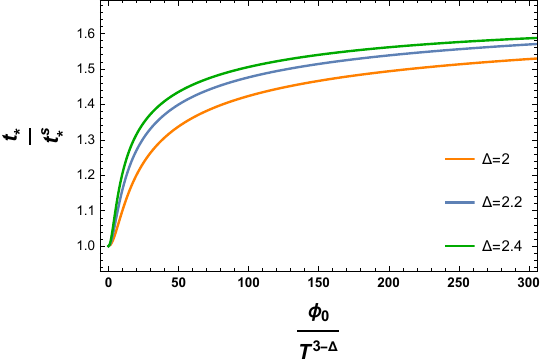}
		\end{subfigure}
		\hspace{0.2cm}
		\begin{subfigure}{0.45\textwidth}
			\centering
			\includegraphics[trim={0 -0.615cm 0 0},clip,width=\textwidth]{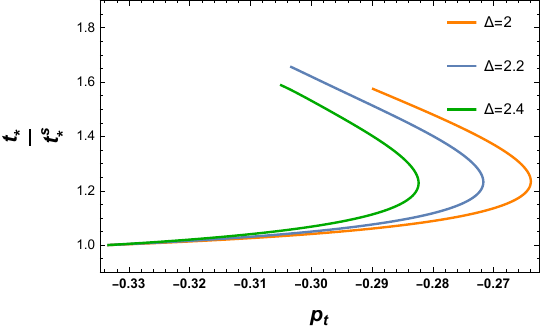}
		\end{subfigure}
  \vspace{-0.4cm}
		\caption{Scrambling time $t_*$ normalized with respect to its Schwarzschild value $t_*^s$ as a function of deformation parameter $\frac{\phi_0}{T^{d-\Delta}}$ (left) or as a function of $p_t$ (right) for $d=2$ (top) and $d=3$ (bottom), and different values of $\Delta$.}
		\label{fig:scrambling}
\end{figure}

With the above equation at hand, we can now investigate the behavior of the scrambling time in the holographic RG flows in consideration. In Fig.~\ref{fig:scrambling} we plot the results that we obtain for the scrambling time as a function of the strength of the deformation $\frac{\phi_0}{T^{d-\Delta}}$ for $d=2$ and $d=3$ and various values of $\Delta$. As anticipated, we find that the scrambling time increases monotonically as a function of the deformation parameter $\frac{\phi_0}{T^{d-\Delta}}$. We also show the results of $t_*$ vs. $p_t$, to try to understand the dependence of the scrambling on the near-singularity Kasner geometry.

A couple of comments are in order. First,
notice that for large enough deformation, the scrambling time seems to increase without a bound. We cannot verify if $t_*\to\infty$ or if $t_*$ saturates to a constant value since our numerics break down for very large values of $\frac{\phi_0}{T^{d-\Delta}}$. Even so, we believe $\phi_0\gg T^{d-\Delta}$ is an interesting limit to explore, which should be amenable to analytic computation.\footnote{One generally expects $t_*\sim\beta\log{S_{\text{BH}}}\to \infty$ as $\beta\to\infty$ ($T\to0$), since in this limit there is no scrambling of information. Naively, the limit $\phi_0\gg T^{d-\Delta}$ should behave similarly, and our results seem to agree with this expectation. However, we cannot completely rule out a different behavior due to the extra scale set by $\phi_0$.} We leave this exploration for future work. Second, as we can observe, the curves of $t_*$ vs. $p_t$ turn out to be multivalued: for a given value of $p_t$ there can be two possible scrambling times. This means we cannot extract the Kasner exponent unequivocally from the scrambling time. This follows from the simple fact that $p_t$ vs. $\frac{\phi_0}{T^{d-\Delta}}$
is non-monotonic in our RG flows (see Fig.~\ref{ptvsphi}), so the relation between them is non-invertible. Physically, this implies that subleading corrections to the Kasner regime will be needed to fully determine the scrambling time, a property that can be attributed to the black hole's horizon. This should not come as a surprise. It is a well-known fact in holography that the properties of the black hole are not only determined by the leading asymptotic boundary values of the fields (non-normalizable modes) but also by the subleading values (normalizable modes). The same situation should apply if we insist on expressing the same observables in terms of near-singularity data. Since the bulk equations are second order, we would need two conditions to fully characterize a given solution. These may be given near the AdS boundary, as commonly done in holography, or at any other place, for example, in the near-singularity region.

\subsubsection{Butterfly velocity}\label{buen}
A further diagnosis of quantum chaos comes from considering the response of the
system to \emph{local} perturbations, as opposed to homogeneous ones. This can be accomplished by upgrading the OTOC (\ref{OTOCchaos}) to
\begin{equation}
    F(t,\vec{x})=\frac{\braket{{\mathcal{O}}(0){\mathcal{W}}(t,\vec{x}){\mathcal{O}(0)}{\mathcal{W}}(t,\vec{x})}}{\braket{{\mathcal{O}(0)}\mathcal{O}(0)}\braket{{\mathcal{W}}(t,\vec{x})\mathcal{W}(t,\vec{x})}}\,,\label{OTOCchaoslocal}
\end{equation}
where the operators $\mathcal{W}$ are now inserted a particular point in space $\vec{x}$. In this case, (\ref{generalOTOC}) generalizes to \cite{Roberts:2014isa}
\begin{equation}
    F(t,\vec{x})\sim 1-\frac{f_0}{N^2}e^{\lambda_L \left(t-\frac{|\vec{x}|}{v_B}\right)}+\cdots\sim 1-e^{\lambda_L \left(t-t_*-\frac{|\vec{x}|}{v_B}\right)}+\cdots\,,\,,
\end{equation}
where $v_B$ is the so-called butterfly velocity. The butterfly velocity estimates the speed of propagation of a localized perturbation that falls into the black hole. From the boundary theory perspective, this quantity defines an emergent light cone, defined by $t-t_*>|\vec{x}|/v_B$. Within the cone, one has that $C(t,\vec{x})=-\langle [\mathcal{W}(t,\vec{x}),\mathcal{O}(0)]^2\rangle\sim O(1)$ while outside the cone, one has $C(t,\vec{x})\approx 0$. Thus, $v_B$ acts as a Lieb-Robinson velocity, setting a
bound for the rate of transfer of quantum information \cite{Roberts:2016wdl}. 

For planar black holes in AdS, the butterfly velocity has already been computed in a number of models \cite{Blake:2016wvh,Blake:2016sud, Reynolds:2016pmi, Lucas:2016yfl,DiNunno:2021eyf}. Notably, in \cite{Mezei:2016zxg} it was shown that for
asymptotically AdS black holes in Einstein gravity, with matter satisfying the null
energy condition (NEC), the butterfly velocity is upper bounded by
\begin{equation}\label{vBbound}
v_B=v_B^{\text{Sch}}\leq \sqrt{\frac{d}{2(d-1)}}\,,
\end{equation}
where $v_B^{\text{Sch}}$ is the value of the $v_B$ for pure AdS-Schwarzschild (no matter). However, there are known holographic systems that violate the bound: i) certain
RG flows that break explicitly the symmetries of AdS
\cite{Giataganas:2017koz,Gursoy:2018ydr,Gursoy:2020kjd}, and holographic theories without a UV fixed point (non-AdS back holes) \cite{Huang:2016izp,Fischler:2018kwt,Eccles:2021zum}.\footnote{Higher curvature gravities also violate the bound for large values of the couplings \cite{Alishahiha:2016cjk}. However, these theories violate causality unless one includes an infinite tower of higher derivative terms \cite{Camanho:2009vw,Camanho:2014apa}.}

Our RG flows do respect the symmetries of AdS, since the relevant deformation is introduced homogenously through space. Thus, we expect (\ref{vBbound}) to be respected. It is however interesting to study the dependence of $v_B$ with respect to $\frac{\phi_0}{T^{d-\Delta}}$ and determine whether this quantity can offer new insights on the black hole interiors.

There are a number of ways to extract this observable: a shock wave calculation with spatial dependence, analogous to the calculation we did for the scrambling time \cite{Roberts:2014isa}, from entanglement
wedge subregion duality, proposed originally in \cite{Mezei:2016wfz}, or by a pole-skipping analysis \cite{Grozdanov:2017ajz}. We will proceed with the shock wave calculation.

Once again, the starting point is a black hole of the form (\ref{metAnsatz}). In terms of Kruskal coordinates (\ref{Kcoords}) the metric can be written as follows:
\begin{equation}
    ds^2=\left(A(uv)dudv+B(uv)d\vec{x}^2\right)\,,
\end{equation}
where
\begin{equation}
A(uv)=\frac{\beta^2}{4\pi^2r^2}\frac{f(r)e^{-\chi(r)}}{uv}\,,\qquad B(uv)=\frac{1}{r^2}\,.
\end{equation}
The black hole horizon is at $r=r_h$ or $uv=0$. We now perturb this black hole geometry with a \emph{localized} shock sent from the left asymptotic boundary. The shock wave propagates along the $u=0$ surface. For large $t_w \gg \beta$, the stress-energy tensor of the shock is exponentially boosted,
\begin{equation}
    T_{uu}^{\text{shock}}=E e^{\frac{2\pi t_w}{\beta}}\delta(u)\delta(\vec{x})\,.
\end{equation}
This local perturbation transforms the background geometry into,
\begin{equation}
    ds^2=\left(A(uv)dudv+B(uv)d\vec{x}^2+A(uv)\delta(u)\alpha(x)du^2\right)\,.
\end{equation}
Particularly, the shock wave shifts the $v$ coordinate by a space-dependent function. $v\to v+\alpha(x)$. Our goal is to determine the form of this function.

In the presence of the shock wave, Einstein's equations become,
\begin{equation}
G_{\mu\nu} - \frac{d(d-1)}{2}g_{\mu\nu} =T_{\mu\nu}^{\phi}+T_{\mu\nu}^{\text{shock}}\,,
\label{EinsteinMM}
\end{equation}
where $T_{\mu\nu}^{\phi}$ is just the right side of (\ref{EinsteinM}). By expanding $A(u v)$ and $B(uv)$ around the horizon and then replacing the result in (\ref{EinsteinMM}), we find the following equation
\begin{equation}
  (-\partial_i^2+\mu^2)\alpha(x)=\frac{16\pi G_N B(0)}{A(0)}E e^{\frac{2\pi t_w}{\beta}}\delta(\vec{x})\,,  
\end{equation}
where $\mu=\sqrt{\frac{2\pi(d-1)T e^{\chi_h/2}}{r_h}}$ is known as the screening length. The solution for large $|\vec{x}|\gg \mu^{-1}$ is given by,
\begin{equation}
    \alpha(x)=\frac{e^{\frac{{2\pi}}{{\beta}}(t_w-t_*)-\mu|x|}}{|x|^{(d-2)/2}}\,,
    \label{buteq}
\end{equation}
where $t_*$ is the scrambling time, $t_*\sim \frac{\beta}{2\pi}\log(a/G_N)$ with $a$ being a constant. Since the leading correction to $F(t,\vec{x})$ is proportional to the shift $\alpha$, we can immediately be read off the butterfly velocity from (\ref{buteq}),
\begin{equation}
    v_B=\frac{2\pi}{\beta \mu}=\sqrt{\frac{2\pi T r_h e^{-\chi_h/2}}{d-1}}\,.
    \label{butterfly}
\end{equation}
\begin{figure}
		\centering
   \begin{subfigure}{0.45\textwidth}
			\centering
			\includegraphics[trim={0 0 0 0},clip,width=\textwidth]{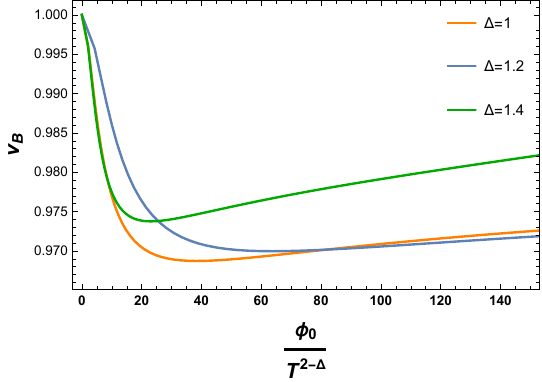}
		\end{subfigure}
		\hspace{0.2cm}
		\begin{subfigure}{0.45\textwidth}
			\centering
			\includegraphics[trim={0 -0.615cm 0 0},clip,width=\textwidth]{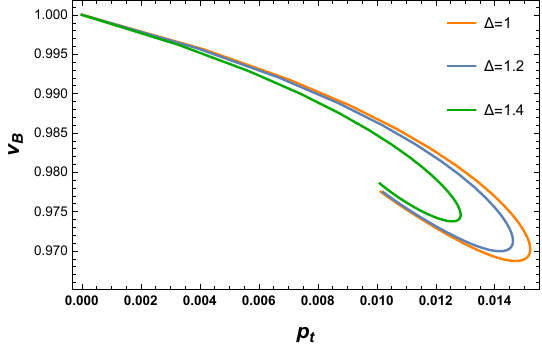}
		\end{subfigure}
		\begin{subfigure}{0.45\textwidth}
			\centering
			\includegraphics[trim={0 0 0 0},clip,width=\textwidth]{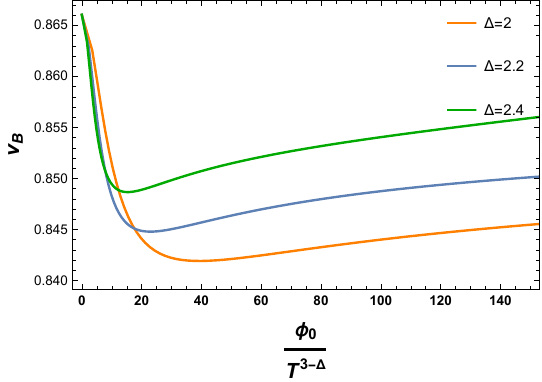}
		\end{subfigure}
		\hspace{0.2cm}
		\begin{subfigure}{0.45\textwidth}
			\centering
			\includegraphics[trim={0 -0.615cm 0 0},clip,width=\textwidth]{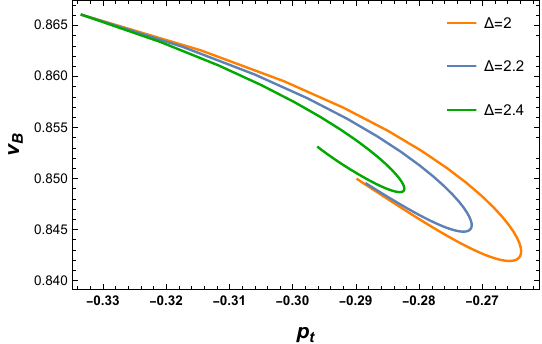}
		\end{subfigure}
  \vspace{-0.4cm}
		\caption{Butterfly velocity as a function of deformation parameter $\frac{\phi_0}{T^{d-\Delta}}$ (left) or as a function of $p_t$ (right) for $d=2$ (top) and $d=3$ (bottom), and different values of $\Delta$.}
		\label{butfig}
\end{figure}

In Fig.~\ref{butfig} we show the results for the butterfly velocity as a function of the deformation parameter $\frac{\phi_0}{T^{d-\Delta}}$ and the Kasner exponent $p_t$. Contrary to the scrambling time, the butterfly velocity displays a non-monotonic behavior as we vary the strength of the deformation. Although our numerics break down for large enough $\frac{\phi_0}{T^{d-\Delta}}$, it seems likely that $v_B$ returns to the original value (without deformation) as $\frac{\phi_0}{T^{d-\Delta}}\to \infty$. It would be interesting to understand this limit better. Other than that, the butterfly velocities do respect the bound (\ref{vBbound}). The plots of $v_B$ vs. $p_t$ are multivalued, as was found for the scrambling time. The same conclusion applies here: it seems likely to us, that one needs subleading terms near the singularity to fully determine observables like $t_*$ and $v_B$ without making reference to the boundary CFT.

\subsubsection{Can OTOCs diagnose the Kasner singularity?\label{sec:hybrid}}

We have seen that observables such as $t_*$ and $v_B$ encode certain information of the RG flow, but are unable to uniquely determine the Kasner exponent $p_t$ that characterizes the region near the singularity. We believe we understand the reason. In the limit of the OTOC relevant for the calculation of $F(t,\vec{x})$ (\ref{OTOCchaoslocal}), the operators $\mathcal{O}$ are inserted at $t_L=t_R=0$. In the heavy-heavy-light-light limit of the OTOC, this implies that geodesic probing of the backreacted geometry intersects the shock wave very near the bifurcate horizon (or, exactly at the bifurcate horizon in the limit $t_w\to\infty$). Meanwhile, the leading correction to geodesic length, and thus to $F(t,\vec{x})$, is proportional to the shift $\alpha$, and comes from a local effect near the intersection point. As a result, both  $t_*$ and $v_B$ can be understood as properties of the near-horizon region of the black hole. This is a well-known fact that can be further explained by recasting the calculation of $F(t,\vec{x})$ in terms of a scattering problem near the black hole horizon \cite{Shenker:2014cwa}.

\begin{figure}
		\centering
		\begin{subfigure}{0.40\textwidth}
			\centering
			\includegraphics[width=\textwidth]{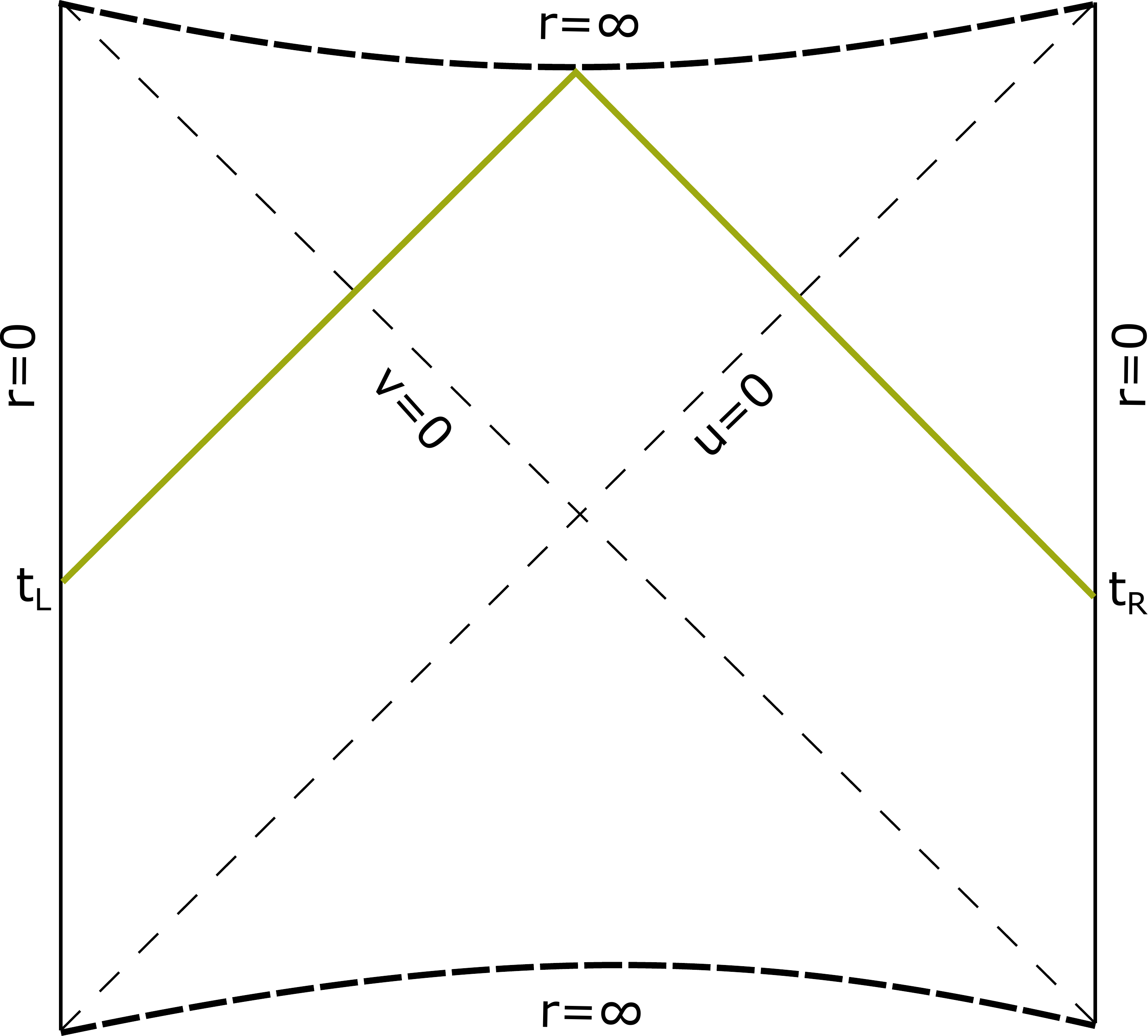}
		\end{subfigure}
		\hfill
		\begin{subfigure}{0.46\textwidth}
			\centering
			\includegraphics[width=\textwidth]{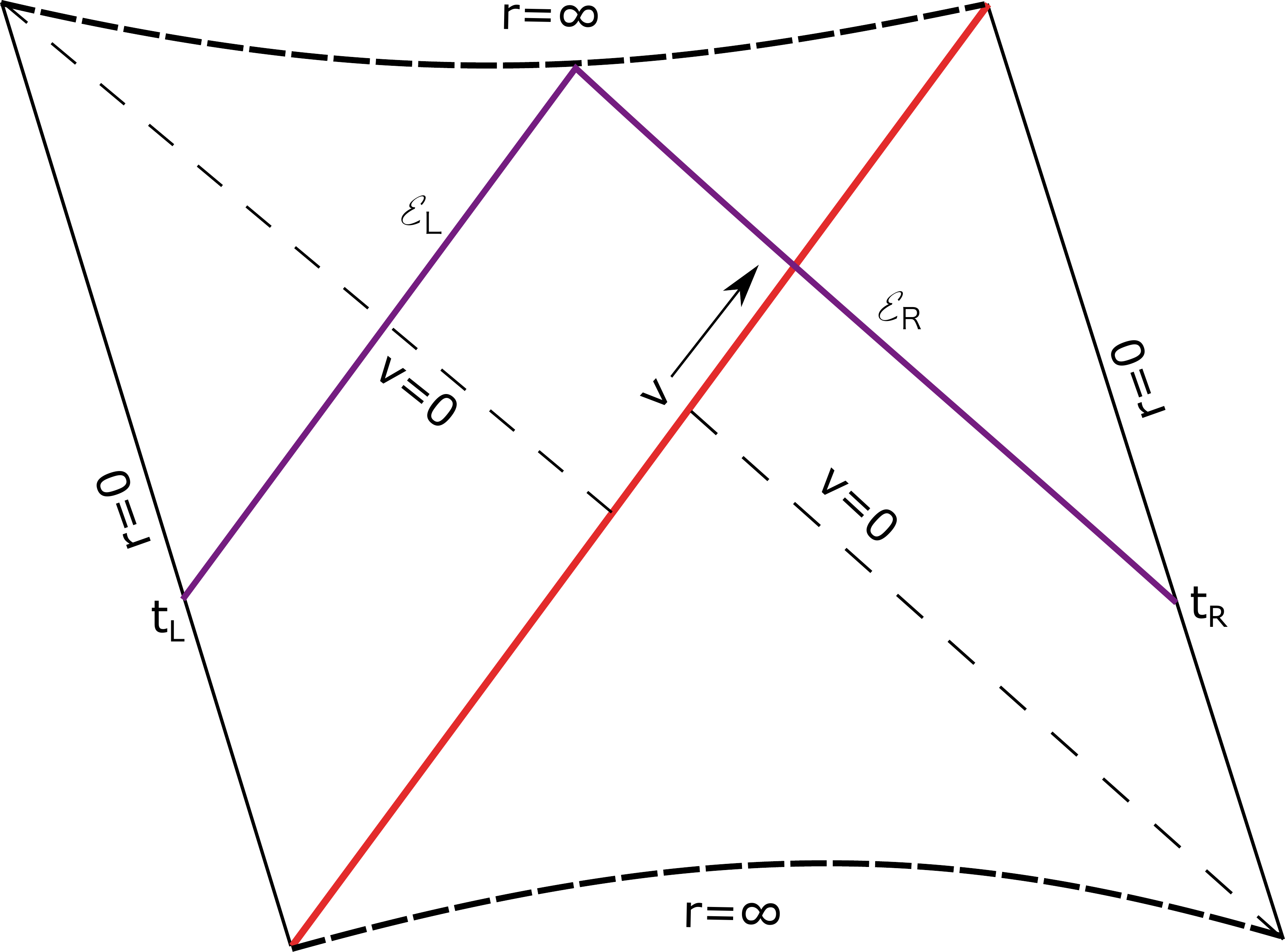}
		\end{subfigure}
		\caption{Null limit of the two-point function in the geodesic approximation (left), and null limit of the OTOC, in the heavy-heavy-light-light approximation (right).}
		\label{fig:null}
\end{figure}

What happens if we now let $t_L$ and $t_R$ to be arbitrary? If we set $t_R=t_L=t_b$ and vary $t_b$ the geodesic will start exploring part of the interior geometry, and eventually, probe the region near the singularity.\footnote{This is not true if $d=2$. In that case $p_t>0$ and this implies that $r_t$ approaches a constant value as $\mathcal{E}\to\infty$ \cite{Hartnoll:2020rwq}. Therefore, we will specialize to $d\geq3$ for the rest of this subsection.} To understand this point, let us analyze first the case without a shock wave, studied briefly in section \ref{sec:geo}. First, note that as we increase the energy of the geodesic, the turning point gets closer to the singularity. Hence, we can use the Kasner metric to determine the turning points for large enough energies,
\begin{equation}
    r_t=\left(\frac{\mathcal{E}^2}{f_1 e^{{-\chi_1}}}\right)^{-\frac{2}{c^2 (d-1)-2 d+4}}+...\,\,\,\,\,\,\,\,\text{as}\,\,\,\,\,\,\,\, \mathcal{E}\to\infty\,.
\end{equation}
In fact, in the strict limit $\mathcal{E}\to\infty$ the geodesic becomes null, reaching the singularity for some finite boundary time $t_b=t_\text{sing}$ \cite{Frenkel:2020ysx}. See Fig.~\ref{fig:null} (left) for an illustration. From (\ref{emptylength}) and (\ref{boundarytime}), it then follows that,
\begin{eqnarray}
    L&=&2\log\bigg(\frac{2}{\mathcal{E}}\bigg)+\frac{\ell_1}{\mathcal{E}}+\cdots+\tilde{\ell}_1 \mathcal{E}^{\frac{c^2 (d-1)+2 d}{c^2 (d-1)-2 d+4}}\,,\\
    t_b&=&t_\text{sing}+\frac{\tau_1}{\mathcal{E}}+\cdots+\tilde{\tau}_1{\mathcal{E}^{\frac{4 (d-1)}{c^2 (d-1)-2 d+4}}}\,.
\end{eqnarray}
The first terms in these expressions come from near-boundary contributions, while the last term corresponds to the leading order contribution near the singularity. Combining these two expressions, we then find the relation,
\begin{equation}
    L(t_b)=2\log(2\Delta t)+c_1 \Delta t+c_2 (\Delta t)^2+\cdots+\tilde{c}_1 (\Delta t)^{-1/p_t}\,,
\end{equation}
where $\Delta t=|t_b-t_\text{sing}|$. This formula is valid for arbitrary dimensions $d\geq 3$ and thus generalizes the result of \cite{Frenkel:2020ysx}. Note that the term coming from the singularity is generally non-analytic, while those coming from the boundary are all analytic. This means that, studying the non-analytic corrections to two-point correlation functions in the limit $t\to t_{\text{sing}}$ is in principle sufficient to extract the Kasner exponent $p_t$ and thus recover the near-singularity geometry.

Let us now come back to the case with a shock wave. In this scenario, we also find that the turning point reaches the singularity as $\mathcal{E}_L\to\infty$, in which case the geodesic becomes null. See Fig.~\ref{fig:null} (right). Thus, on general grounds, we expect the same type of contribution to the length coming from the near-singularity region, $L\sim (\Delta t)^{-1/p_t}$, perhaps with a shift to $t_{\text{sing}}$ of order $\alpha$. Naively, this would imply that the four-point OTOC should also suffice to diagnose $p_t$, by studying certain time limits. In particular, upon analytic continuation, we would expect that the relevant OTOC should be of the form $\langle\mathcal{O}(t_b)\mathcal{W}(t_w)\mathcal{O}(-t_b)\mathcal{W}(t_w)\rangle$, with $t_w\gg\beta$ and $t_b\approx t_{\text{sing}}$. However, this naive expectation is not true. It was recently shown that in this limit, the null geodesic is not the relevant saddle for the position-space correlator \cite{Horowitz:2023ury}, which is instead given by a complex geodesic. Nevertheless, \cite{Horowitz:2023ury} showed that the null limit does feature in a `hybrid' correlator where one considers the Fourier transform of $t_L$ and takes the limit $\omega_L\to\infty$. Based on their result, we can expect that the peculiar non-analytic contribution to the length from the near-singularity does show up in a `hybrid' OTOC
\begin{equation}
\langle\mathcal{O}(t_b)\mathcal{W}(t_w)\mathcal{O}(-\omega_b)\mathcal{W}(t_w)\rangle\sim e^{-\Delta \tilde{c_1}|t_b-t_{\text{sing}}|^{-1/p_t}+\cdots}\,,\label{OTOTpt}
\end{equation}
in the limit $t_w\to\infty$, $\omega_b\to\infty$, $t_b\approx t_{\text{sing}}$. It would be interesting to understand this result from a field theory calculation. Note that in this case, the intersection between the shockwave and the geodesic still happens at the horizon, but far from the bifurcation point. However, one could engineer a situation in which the two intersect very close to the singularity, e.g., by varying $t_w$. Perhaps, in such a scenario, one could also try to understand (\ref{OTOTpt}) by reformulating the bulk calculation of the OTOC in terms of the scattering of quanta close to the singularity, analogous to the calculation of \cite{Shenker:2014cwa}. We leave these explorations for future work.

\subsection{Thermal $a$-function}
In this section, we will explore the characteristics of the thermal $a$-function, first introduced in \cite{Caceres:2022smh} and later expanded upon in \cite{Caceres:2022hei}. Our focus will be on the effect of a shock wave sent from the left asymptotic boundary at some arbitrary boundary time $t_w$. Before doing that, let us briefly discuss the holographic `trans-IR' flows that are accessible via the black hole interiors.

The core concept of a `trans-IR' flow involves an analytic continuation of the conventional RG flow beyond its infrared (IR) fixed point to complex energies. The $a$-function is monotonic along the entire RG flow, including both the conventional RG flow, defined outside the horizon, and the `trans-IR' portion of the flow, defined inside the black hole, i.e., the `cosmological interior.' To define this function, we start with the following black hole metric,
 \begin{equation}
ds^2 = e^{2A(z)}\left[-h(z)^2 dt^2 + d\vec{x}^2\right] + dz^2,
\end{equation}
with $t \in \mathbb{R}$, $\vec{x} \in \mathbb{R}^{d-1}$, $z \geq 0$. These are known as domain-wall coordinates. Holographically, $z$ represents an energy scale in the theory. We assume that $h(z)$ has a simple root at the black hole horizon $z=0$, i.e., the IR. The AdS boundary is located at $z \to \infty$ and it therefore corresponds to the UV of the theory. With this metric, the thermal $a$-function is defined as \cite{Caceres:2022smh},
\begin{equation}
a_T(z) = \frac{\pi^{d/2}}{\Gamma\left(\frac{d}{2}\right)\ell_P^{d-1}}\left[\frac{h(z)}{A'(z)}\right]^{d-1}.\label{aFunction0}
\end{equation}
It is straightforward to show that this function is monotonic. This can be achieved by demanding that the matter fields respect the null energy condition (NEC), $T_{\mu\nu}k^\mu k^\nu\geq 0$, where $T_{\mu\nu}$ is the stress-energy tensor and $k^\mu$ is an arbitrary null vector. Picking the radial null vector $k^{\mu}=e^{-A(z)}\delta^{\mu}_t+\delta^{\mu}_{z}$, and using the bulk Einstein's equations, one can show the monotonicity of the thermal $a$-function by evaluating,
\begin{equation}
    \frac{d a_T}{d z}\sim \frac{1}{A'(z)^d}\bigg(T^{z}_{z}-T^t_t\bigg)\geq 0\,.
\end{equation}
It is also easy to check that this function is stationary at the AdS boundary and at the horizon, i.e., $\frac{d a_T}{dz}|_{z\to\infty}=\frac{d a_T}{dz}|_{z=0}=0$ \cite{Caceres:2022smh}. These are fixed points of the RG flow.

Note that the above coordinates are only defined in the exterior of the black hole. However, we can access the black hole interior by the following analytic continuation of the time and radial coordinate,
\begin{equation}
    t=t_I-\text{sgn}(t_I)\frac{i \gamma}{2\mathcal{T}}\,,\qquad z=i\kappa\,,
\end{equation}
where $\gamma$ is some half integer and $\mathcal{T}\equiv\frac{e^{A(0)}h'(0)}{2\pi}$.
To demonstrate monotonicity inside, it is useful to employ the coordinate patch from (\ref{metAnsatz}), which can access the black hole interior without the need for analytic continuation. We make a coordinate transformation, $z\to z(r)$, and some identifications to find the $a$-function in this patch. This transformation amounts to:
\begin{eqnarray}
    \frac{d z}{dr}&=&-\frac{1}{r\sqrt{f(r)}}\,,\label{transzr}\\
    e^{2 A(z)}&=&\frac{1}{r^2}\label{coordtrans0}\,,\\
    h(z)^2&=&f(r)e^{-\chi(r)}\,.
    \label{coordtran01}
\end{eqnarray}
By substituting (\ref{coordtrans0})-(\ref{coordtran01}) in (\ref{aFunction0}), we finally get,
 \begin{equation}
a_T(r) = \frac{\pi^{d/2}}{\Gamma\left(\frac{d}{2}\right)\ell_P^{d-1}} e^{-(d-1)\chi(r)/2}\,.\label{aFunctionSch}
\end{equation}
Further, by using Einstein's equations, we find that,
\begin{equation}
    \frac{d a_T}{d r}\sim -f(r)\bigg(T^r_r-T^t_t\bigg)\leq 0\,,\label{derivar}
\end{equation}
thus proving that the $a$-function is also monotonic inside of the black hole, i.e., for $r>r_h$.\footnote{Since $\partial_r$ becomes time-like for $r>r_h$ it is interesting to note that $a_T(r)$ can be used to define a relational notion of a `clock' in the `cosmological interior,' given its monotonicity (or rather, $1/a_T(r)$, which is an increasing function). Such cosmologies can in principle be solved for given consistent initial data by considering the Wheeler-DeWitt equation for the black hole interior \cite{Hartnoll:2022snh}.} In summary, the $a$-function is monotonic throughout the entire flow and becomes stationary at the fixed points.\footnote{Note that, because of the coordinate transformation (\ref{transzr}), $d a_T/dr$ is not necessarily zero at the horizon. However, if we insist that $z$ should map to the physical energy $E$, we can identify the horizon as a fixed point of the RG flow. Other points where $d a_T/dr=0$ should also imply $d a_T/dz=0$ and viceversa, provided $dr/dz\neq\{0,\infty\}$.} These properties allow us to establish a connection between this monotonic function and the total number of degrees of freedom as we vary the energy scale in the corresponding dual field theory \cite{Caceres:2022smh}.

\begin{figure}
    \centering
\centering
		\begin{subfigure}{0.45\textwidth}
			\centering
			\includegraphics[width=\textwidth]{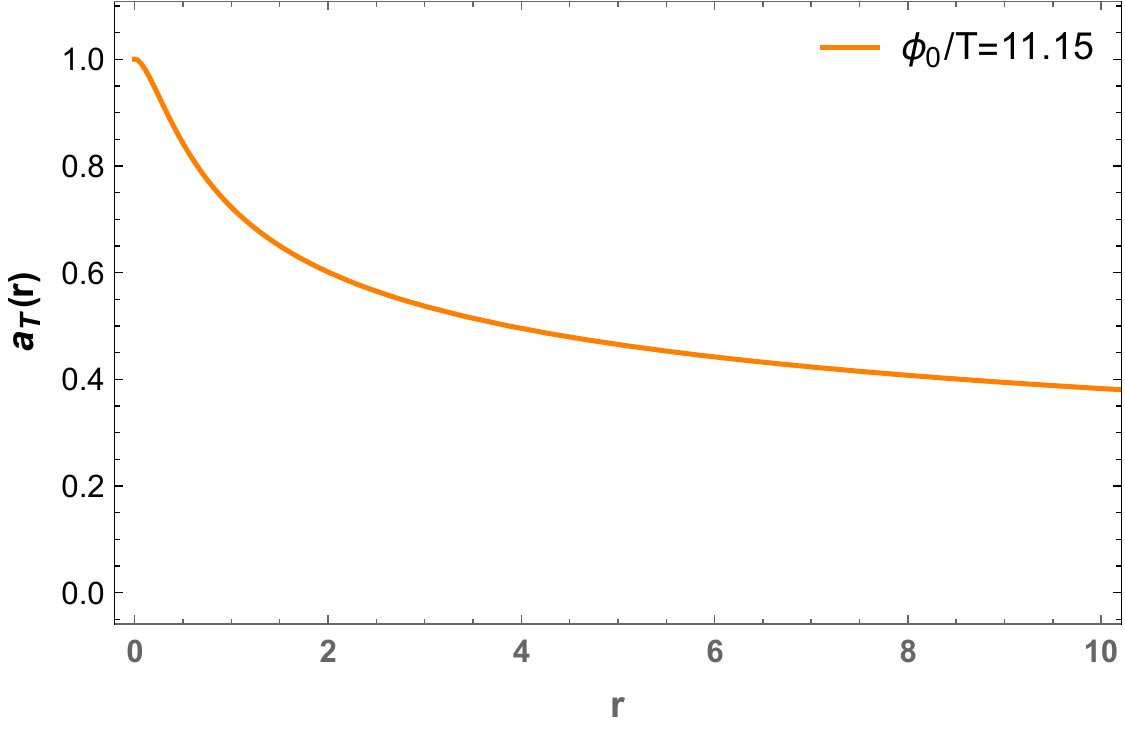}
		\end{subfigure}
		\hfill
		\begin{subfigure}{0.45\textwidth}
			\centering
			\includegraphics[width=\textwidth]{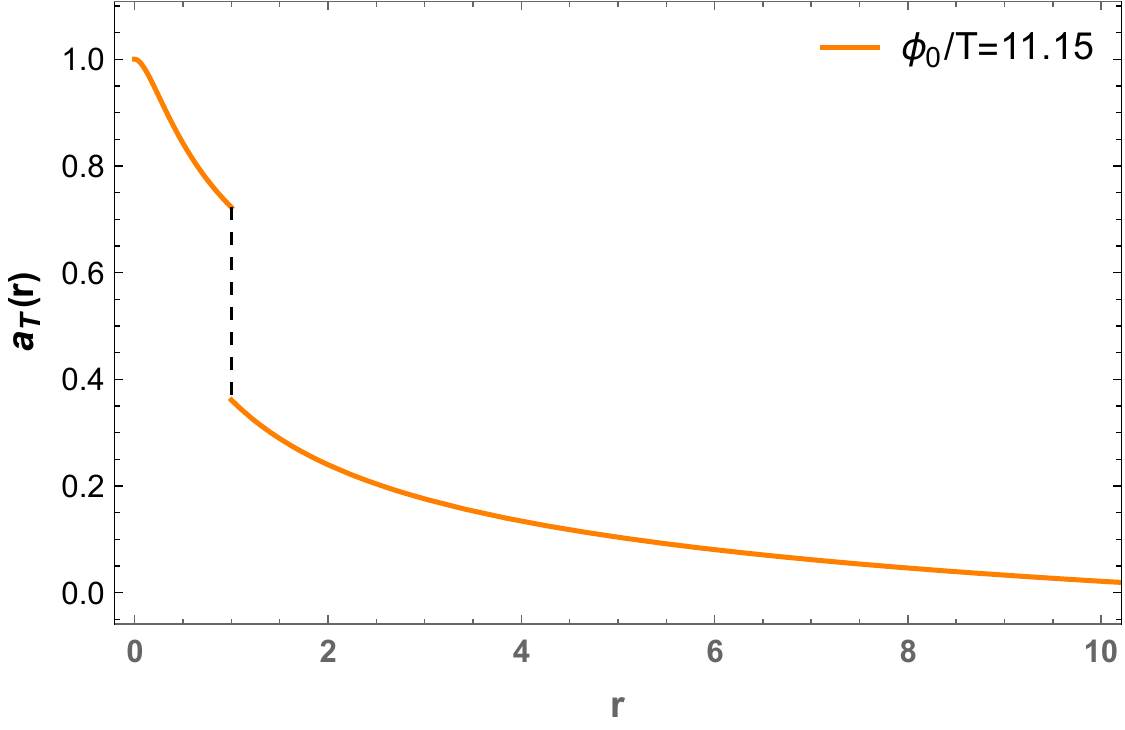}
		\end{subfigure}

    \caption{Thermal $a$-function in the unperturbed RG flow (left) and in the presence of a shock wave sent at $t_w\to\infty$ (right). The discontinuity happens exactly at the horizon $r=r_h$, and its magnitude is of order $O(E/M)$ for $E\ll M$.}
    \label{dipafunction}
\end{figure}

\begin{figure}
		\centering
            \includegraphics[width=0.6\textwidth]{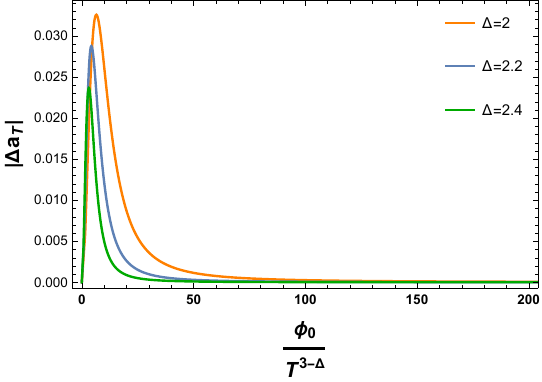}
		\caption{Discontinuity in the thermal $a$-function across a shock wave sent at $t_w\to\infty$ as a function of deformation parameter $\frac{\phi_0}{T^{d-\Delta}}$, for $d=3$ and various values of $\Delta$.}
		\label{afunctionasafunctionofphi}
\end{figure}

With this introduction, let us now investigate any possible imprints of a shock wave on the thermal $a$-function of our holographic RG flows. To begin with, it is clear that $a_T(r)$ has knowledge of the full black hole interior. In fact, from its behavior at $r\to\infty$ we can already extract the information about the Kasner singularity \cite{Caceres:2022smh}:
\begin{equation}
a_T(r)\sim C_d\, r^{-\frac{(d-1)[d(1+p_t)-2]}{(1-p_t)}}\quad\text{as}\quad r\to\infty\,,\label{aTkasner}
\end{equation}
where $C_d>0$ is a constant. This relation should still be the same in the presence of a shock wave, with a small change in $p_t$ given the change in the mass of the black hole, hence the change in the temperature $T$ and the ratio $\frac{\phi_0}{T^{d-\Delta}}$. However, there is a more drastic effect on  $a_T(r)$. More specifically, we observe that the $a$-function undergoes a finite discontinuity at the location of the shock wave. This discontinuity follows from the fact that the derivative of the $a$-function is proportional to the trace of the stress-energy tensor (\ref{derivar}). When a shock wave is introduced, it results in a delta function in the stress-energy tensor, which leads to a finite discontinuity in the $a$-function. When the shock wave is sent at the infinite past ($t_w\to \infty$), we can compute the discontinuity analytically in terms of near-horizon data. More specifically, by evaluating the function (\ref{aFunctionSch}) in the two different patches glued at the horizon we find that, at $O(E/M)$:
\begin{equation}
    \Delta{a_T}=\frac{\pi^{d/2}}{\Gamma\left(\frac{d}{2}\right)\ell_P^{d-1}}\bigg(e^{-(d-1)\tilde{\chi}(r_h)/2}-e^{-(d-1)\chi(r_h)/2}\bigg)=-\frac{E\pi^{d/2}}{\Gamma\left(\frac{d}{2}\right)\ell_P^{d-1}}\frac{dr_h}{dM}A(r_h,r_h)\,,
    \label{atdiscon}
\end{equation}
where $\chi(r)$ and $\tilde{\chi}(r)$ are `pre' and `post' shock wave metric functions and
$A(r_h,r_h)$ is determined from the expansion of the latter near the horizon, $e^{-(d-1)\tilde{\chi}(r)/2}= \text{const.} + (r-r_h)A(r,r_h)+O[(r-r_h)^2]$. In Fig.~\ref{dipafunction} we plot our results for the thermal $a$-function as a function of $r$ and contrast it with respect to the case without a shock wave. Physically, the shift in $a_T$ implies that the positive energy shock removes some degrees of freedom at the horizon, leading to a sudden drop in the thermal $a$-function. The discontinuity happens in this case exactly at the horizon and its magnitude (\ref{atdiscon}) is of order $O(E/M)$, as we are considering the limit $E\ll M$. We further plot $|\Delta{a}_T|$ as a function of the deformation parameter $\frac{\phi_0}{T^{d-\Delta}}$ in Fig.~\ref{afunctionasafunctionofphi}. 

\begin{figure}
		\centering
		\begin{subfigure}{0.4\textwidth}
			\centering
			\includegraphics[width=\textwidth]{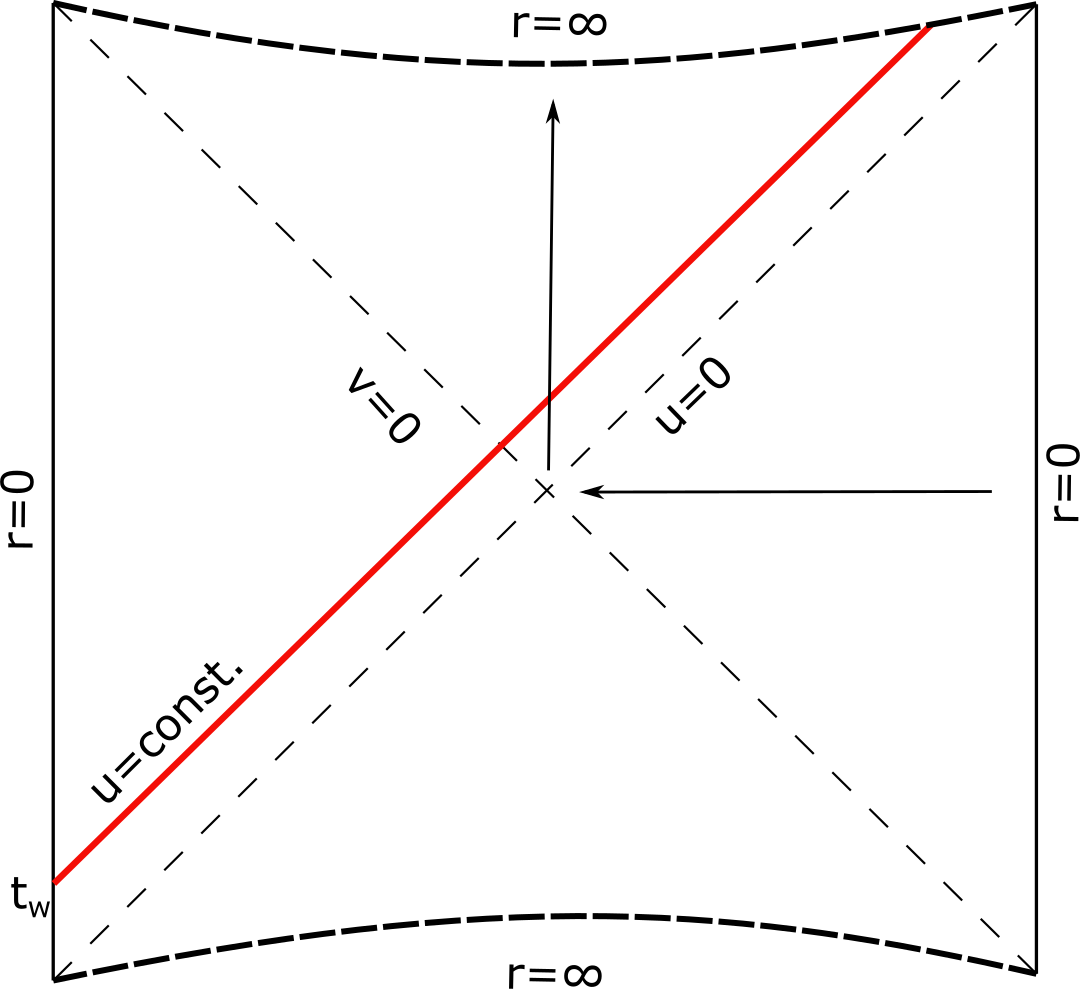}
		\end{subfigure}
		\hfill
		\begin{subfigure}{0.54\textwidth}
			\centering
			\includegraphics[width=\textwidth]{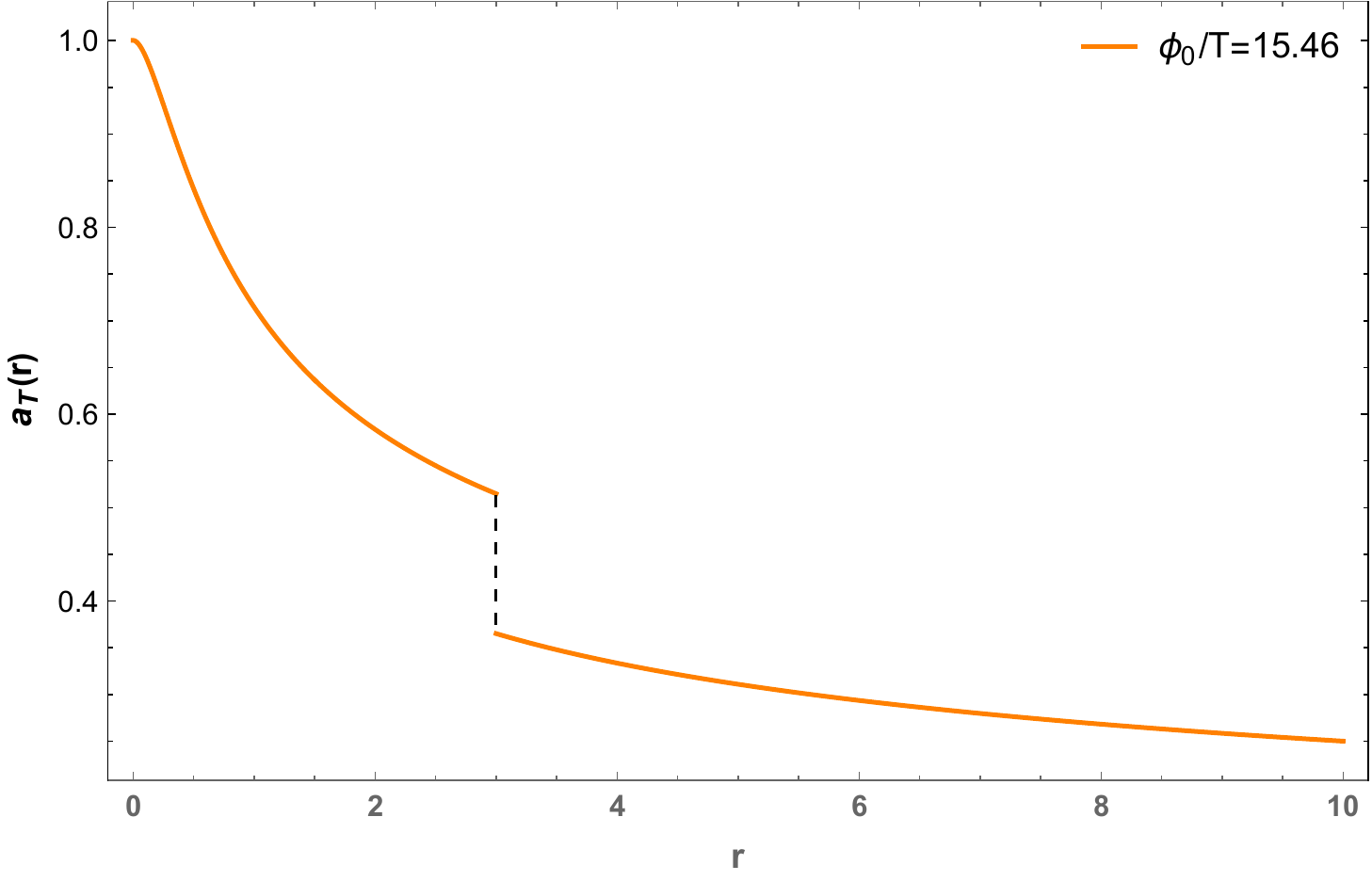}
		\end{subfigure}
		\caption{A shock wave is sent from the left asymptotic boundary at $t_w$ (left). Because of this perturbation, the thermal $a$-function changes discontinuously across the shock wave (right). The point of discontinuity is determined by the position of the shock wave at the time of the measurement. In this example, we have picked $t=0$, which yields $r=3$. In contrast, the black hole horizon is located at $r_h=1.5$.}
		\label{aundersf}
\end{figure}

We can also analyze the case where the shock wave is sent at some arbitrary $t_w$. Qualitatively, a similar effect is observed, i.e., a sudden jump in the thermal $a$-function, given by a formula similar to (\ref{atdiscon}). However, the discontinuity does not happen at the horizon in the more general case. The exact point of the transition can be determined by finding the position of the shock $r_{\text{shock}}$, which depends on the time $t$ we chose to perform the measurement. This should not come as a surprise. We recall that the shock wave geometry can be interpreted as a `quenched'
state in the theory which can be obtained by a time-dependent Hamiltonian. Thus it makes sense that the thermal $a$-function, which measures the number of degrees of freedom at a given energy scale, could be time-dependent as well. In Fig.~\ref{aundersf} we show the results for the $a$-function in a simple example, where we have fixed the time of the measurement $t$. The transition point happens at some $r_{\text{shock}}>r_h$. Since the region $r>r_h$ codifies the evolution of the cosmological interior,
the jump can also be interpreted as a `cosmological time skip.'\footnote{In the $t_w\to\infty$ case, the jump merely shifts the initial time of the cosmological evolution.} To understand this effect, we must think in terms of an observer falling into the singularity and encountering the shock wave along its trajectory. The shock wave deforms the geometry and, as a result, induces a relativistic effect in the observer that is perceived as an infinitely boosted length contraction. Since $\partial_r$ is time-like in the interior of the black hole, the length contraction is then interpreted as a time contraction, which explains the term `time skip.' We recall that the bulk matter is assumed to respect energy conditions (the NEC, in particular), so the time jump is always future-directed. That is, we cannot jump back in time unless we violate the assumed energy conditions.

\subsection{Entanglement velocity}

Another observable that is sensitive to the black hole interior is the so-called entanglement velocity \cite{Hartman:2013qma}. In a quenched system, the entanglement entropy of a large subsystem $S_A$ grows as \cite{Liu:2013iza,Liu:2013qca} $d S_A/dt=v_E s_\text{th}\Sigma_A$ for $t_{\text{th}}\ll t \ll t_{\text{sat}}$, where $t_{\text{th}}\sim\beta$ is the local thermalization time and $t_{\text{sat}}$ is the saturation time, which scales with the characteristic size of the system $t_{\text{sat}}\sim L_R$. Here $s_\text{th}$ is the equilibrium entropy density, $\Sigma_A$ is the area bounding the subsystem and $v_E$ is the entanglement velocity. The above relation
was initially derived for a quenched black-hole system with a single boundary. However, \cite{Hartman:2013qma} showed that the same relation applies to the case of a two-sided black hole (TFD state) undergoing the usual Hamiltonian evolution. The map between the two follows by cutting the
usual eternal black hole Penrose diagram in half by adding an end-of-the-world brane in the bulk, producing the so-called `B-states.'

The entanglement velocity $v_E$ has been shown to be upper bounded by the butterfly velocity $v_B$ in general quantum chaotic systems \cite{Mezei:2016zxg},
\begin{equation}
v_E\leq v_B\,,\label{ratiovevb}
\end{equation} 
where they further uncovered some interesting connections between the two. Given that $v_E$ can probe part of the black hole interior, while $v_B$ can be entirely derived from near-horizon physics, for completeness, here we will present the calculation of $v_E$ in the RG flows considered in this paper and contrast the results with those we obtained earlier for the butterfly velocity $v_B$. 

\begin{figure}
		\centering
		\begin{subfigure}{0.45\textwidth}
			\centering
                \includegraphics[width=\textwidth]{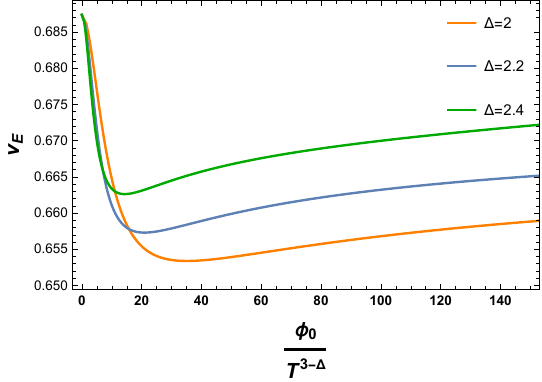}
		\end{subfigure}
		\hfill
		\begin{subfigure}{0.45\textwidth}
			\centering
		\includegraphics[trim={0 -0.615cm 0 0},clip,width=\textwidth]{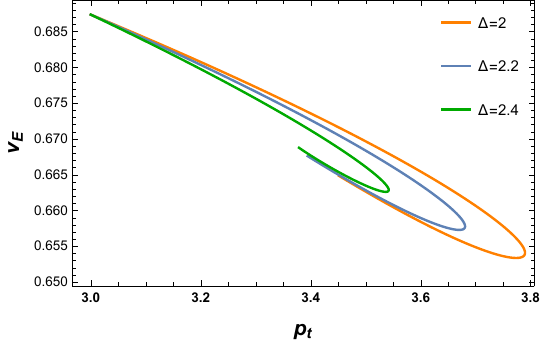}	
		\end{subfigure}
		\caption{Entanglement velocity as a function of the deformation parameter $\frac{\phi_0}{T^{3-\Delta}}$ (left) and the Kasner exponent $p_t$ (right) for $d=3$ and various values of $\Delta$.}
		\label{ev}
\end{figure}

\begin{figure}
		\centering
		\begin{subfigure}{0.45\textwidth}
			\centering
               \includegraphics[width=\textwidth]{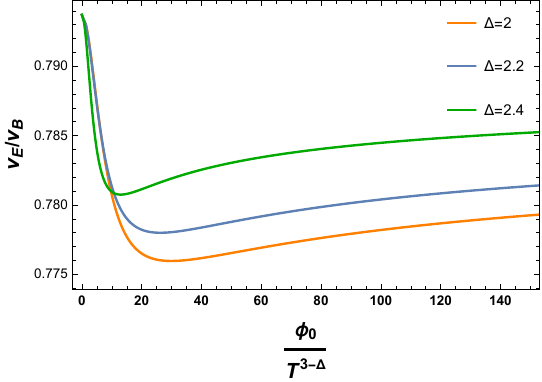}
		\end{subfigure}
		\hfill
		\begin{subfigure}{0.45\textwidth}
			\centering
		\includegraphics[trim={0 -0.615cm 0 0},clip,width=\textwidth]{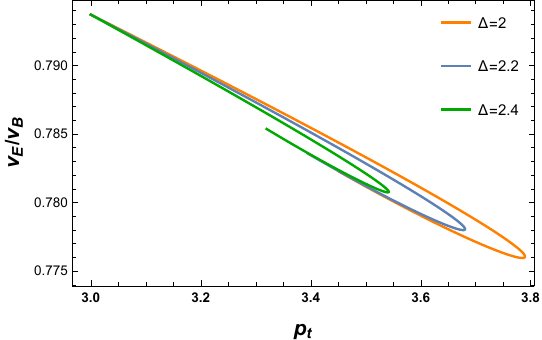}	
		\end{subfigure}
		\caption{Ratio between entanglement velocity and butterfly velocity as a function of the deformation parameter $\frac{\phi_0}{T^{3-\Delta}}$ (left) and the Kasner exponent $p_t$ (right) for $d=3$ and various values of $\Delta$. For any amount of relevant deformation, the entanglement velocity is always less than the butterfly velocity, as expected.}
		\label{rvevb}
\end{figure}

To start, we consider symmetric entangling surfaces probing the deformed interiors. This can be accomplished by picking the subsystem to be $A=A_L\cup A_R$, where $A_L$ and $A_R$ are half-spaces on the left and right boundaries, respectively. The induced metric on the entangling surface that corresponds to this subsystem is given by
\begin{equation}
    ds^2_{\text{induced}}=\frac{1}{r^2}\left[\left(-f(r)e^{-\chi(r)}t'(r)^2+\frac{1}{f(r)}\right)dr^2+d\vec{x}_{d-2}^2\right]\,,
\end{equation}
leading to the following area functional:
\begin{equation}
    \mathcal{A}=\int\frac{dr}{r^{d-1}}\sqrt{-f(r)e^{-\chi(r)}t'(r)^2+\frac{1}{f(r)}}=\int dr \mathcal{L}\,.
\end{equation}
Like the symmetric geodesics, the above Lagrangian $\mathcal{L}$ does not explicitly depend on time. Thus, we can define a conserved quantity $\mathcal{E}$ which is constant over the entire spacelike surface,
\begin{equation}
    \mathcal{E}=-\frac{\partial \mathcal{L}}{\partial t'(r)}\implies t_b=-P\int_{0}^{r_t}\frac{\text{sgn}(\mathcal{E})e^{\chi(r)/2}dr}{f(r)\sqrt{1+f(r)e^{-\chi(r)}/{(r^{d-1}\mathcal{E})}^2}}\,,
\end{equation}
where $t_b$ is the boundary anchoring time for the bulk entangling surface, and $r_t$ is the turning point defined by the relation,
\begin{equation}
    \mathcal{E}^2=-\frac{f(r_t)e^{-\chi(r_t)}}{r_t^{2(d-1)}}\,.
\end{equation}
With these expressions at hand, we find the area of the extremal surface to be
\begin{equation}
    \mathcal{A}=2\int_0^{r_t}\frac{dr}{\sqrt{f(r)+e^{\chi(r)}{(r^{d-1}\mathcal{E}^2})}}\,.
    \label{ee}
\end{equation}
At late times ($\mathcal{E}\to\infty$), the extremal surfaces are trapped on a slice with $r_t=r_{\text{crit}}$ inside the horizon. This causes their areas to exhibit linear growth, which indicates a velocity for the associated entanglement entropies. We obtain this rate of growth of the entropy at late times from 
(\ref{ee}),
\begin{equation}
    \frac{\partial \mathcal{S}}{\partial t_b}=\frac{2 v_E}{4 G_N r_{h}^{d-1}}=v_E s_{\text{th}},\,\,\,\,\,\,\,\,\,\,\,\,\,\,\,\,\,v_E^2=r_{h}^{2(d-1)}\frac{|f|e^{-\chi}}{r^{2(d-1)}}\bigg|_{r=r_{\text{crit}}}\,,
\end{equation}
where $\mathcal{S}=\frac{\mathcal{A}}{4 G_N}$. We plot the entanglement velocity $v_E$ as a function of the deformation parameter $\frac{\phi_0}{T^{d-\Delta}}$ and Kasner exponent $p_t$ in Fig.~\ref{ev}.  In general, we observe that the entanglement velocity's dependence on the deformation exhibits a qualitatively similar pattern to what we observed for the butterfly velocity, demonstrating a non-monotonic behavior. Furthermore, although these extremal surfaces probe part of the interior geometry, their dependence on $p_t$ indicates we cannot understand $v_E$ solely as a feature of the near-singularity region. It is likely that subleading corrections near the singularity could completely determine $v_E$, as we previously argued in the case of the butterfly velocity. Finally, we depict the ratio between these two velocities as a function of the deformation parameter and Kasner exponent in Fig.~\ref{rvevb}, which demonstrates that the bound (\ref{ratiovevb}) is respected along the full RG flow. All these conclusions remain valid for any $d\geq2$. However, for brevity, we have only presented plots for the case of $d=3$.

\section{Bouncing interiors}{\label{Bounces}}
Recently, it has been shown that the interiors of planar AdS black holes, which are solutions of the Einstein-Scalar system (\ref{action}) with an even super-exponential potential, exhibit an infinite sequence of Kasner epochs \cite{Hartnoll:2022rdv}, bearing a resemblance to Misner's mixmaster universe \cite{Misner:1969hg}. We will use this model to explore how the different field theory observables are affected by the intricate nature of the black hole interior. 

To begin with, let us briefly review the relevant features of the model. For simplicity, we will fix the number of dimensions to $d=3$ so that the gravity action reads,
\begin{equation}
I = \int d^4x \sqrt{-g}\left(R+6-\frac{1}{2}\nabla^\alpha\phi \nabla_\alpha\phi - V(\phi)\right)\,,\label{actionbounce}
\end{equation}
with a potential given by $V(\phi)=\frac{1}{2}m^2\phi^2+\frac{1}{10}\exp[\frac{\phi^8}{10}]$. We will further consider the following metric ansatz, which is similar to (\ref{metAnsatz}) but specialized to $(3+1)$-dimensions,
\begin{equation}
 ds^2 = \frac{1}{e^{2\rho}}\left[-f(\rho)e^{-\chi(\rho)} dt^2 + \frac{d\rho^2}{e^{-2\rho}f(\rho)} + d{x}^2+d{y}^2\right]\,.
 \label{newmetric}
\end{equation}
Note that we have introduced a new radial coordinate $\rho$, related to $r$ via
\begin{equation}
    r=\exp({\rho})\,.
\end{equation}
We do so in order to improve the accuracy of the numerics for large values of $r$. By plugging this ansatz into the equations of motion, we find
\begin{eqnarray}
    2\phi ''(\rho )f(\rho)+2 f'(\rho ) \phi '(\rho )-f(\rho ) \chi '(\rho )
   \phi '(\rho )-6 f(\rho ) \phi '(\rho )-2 V'(\phi (\rho ))&=&0\,,\\
   4f'(\rho )-f(\rho ) \phi '(\rho )^2-12 f(\rho )-2
   V(\phi (\rho ))&=&0\,,\\
   2\chi '(\rho )- \phi '(\rho )^2&=&0\,.\label{EE:bouncechiphi}
\end{eqnarray}
We expect the behavior of the fields near the boundary region $\rho\to -\infty$ to be:
\begin{equation}
    \phi\sim\phi_0\, e^{(d-\Delta)\rho}\,,\qquad \chi\sim 0\,,\qquad f\sim1\,,
\end{equation}
where $\phi_0$ is the source of the relevant operator inducing the RG flow. We also assume $f(\rho)$ has a simple root at the horizon $\rho=\rho_h$, in which terms we can express the temperature of the black hole as
\begin{equation}
    T=\frac{|f'(\rho_h)|e^{-\frac{\chi_h+{2\rho_h}}{2}}}{4\pi}\,.
\end{equation}
With these boundary conditions, we can numerically solve the equations of motion. We follow a procedure analogous to that outlined in section \ref{RG Kasner}: we first shoot from the horizon to the boundary to relate near-horizon and near-boundary data, and then we shoot from the horizon to the interior of the black hole. 
\begin{figure}
		\centering
		\begin{subfigure}{0.45\textwidth}
			\centering
			\includegraphics[width=\textwidth]{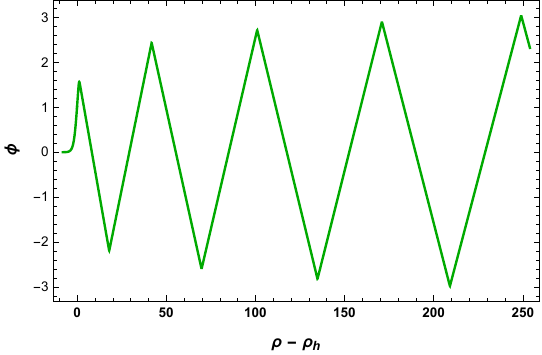}
		\end{subfigure}
		\hfill
		\begin{subfigure}{0.45\textwidth}
			\centering
			\includegraphics[width=\textwidth]{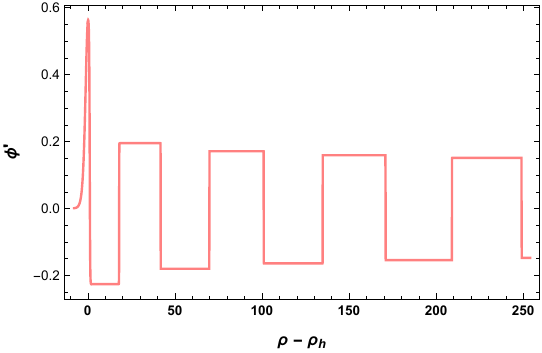}
		\end{subfigure}
		\caption{$\phi$ and ${\phi'}$ as a function of $\rho-\rho_h$ for $\rho_h=1$, $\phi(\rho_h)=1$ and $\Delta=2$. Each region with a linear $\phi$, or constant  $\phi'$, determines a particular Kasner epoch.}
		\label{phiphidotrho}
  \end{figure}

\begin{figure}[t]
		\centering
		\begin{subfigure}{0.48\textwidth}
			\centering
			\includegraphics[width=\textwidth]{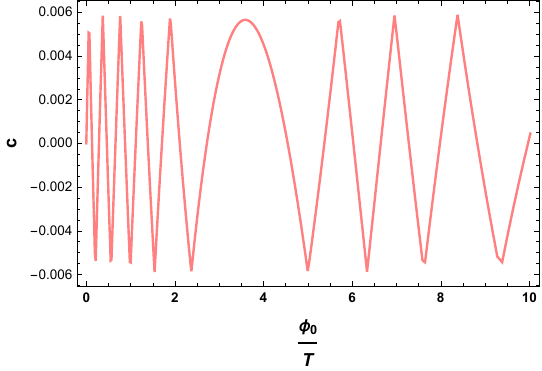}
		\end{subfigure}
		\hfill
		\begin{subfigure}{0.5\textwidth}
			\centering
			\includegraphics[width=\textwidth]{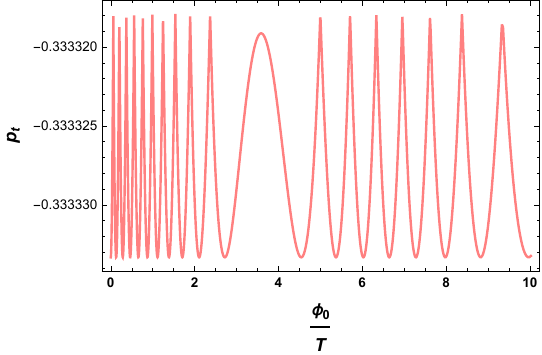}
		\end{subfigure}
		\caption{Behavior of $c$ and $p_t$ vs. $\frac{\phi_0}{T^{3-\Delta}}$ with the super-exponential potential, for $\rho=260$ and $\Delta=2$.}
		\label{ptvsphiTsuper}
  \end{figure}

In Fig.~\ref{phiphidotrho} we plot a typical solution for the scalar field and its derivative as a function of radial coordinate $\rho$, for some sample parameters. A linear behavior in $\phi$, or a constant in $\phi'$, corresponds to an approximate Kasner solution, thus we see that the evolution of the interior undergoes an infinite sequence of Kasner epochs separated by sharp bounces, a feature of the so-called `cosmological billiards' \cite{Damour:2002et}. In each of these epochs, the solution can be approximately described by
\begin{eqnarray}
\phi\sim 2c \rho+\phi_1\,,\qquad\chi\sim 2c^2 \rho+\chi_1\,,\qquad f\sim-f_1 e^{(3+c^2)\rho}+f_2\,,
\end{eqnarray}
where the constant $c$ determines the Kasner exponents through the standard relations (\ref{eq:kasnerc}). The dependence of this constant and the Kasner exponents on $\phi_0/T^{3-\Delta}$ for fixed $\rho$ are highly oscillatory, indicating the chaotic nature of the interiors. We investigate these dependence in Fig.~\ref{ptvsphiTsuper}, for some sample parameters. 
 
In the subsequent sections, we will study the signatures of the bouncing interiors on various observables in the dual field theory.

\subsection{Thermal $a$-function}
\begin{figure}
     \begin{subfigure}{0.42\textwidth}
			\includegraphics[trim={0 -0.15cm 0 0},clip,height=4.4cm]{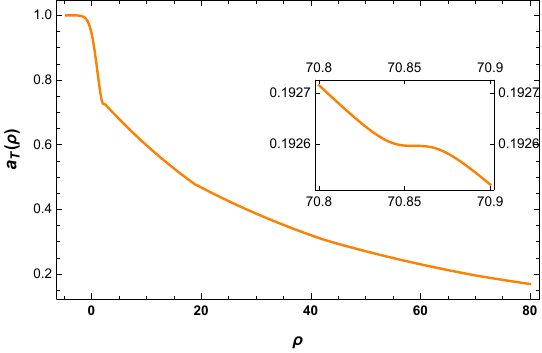}
		\end{subfigure}
		\hspace{1cm}
		\begin{subfigure}{0.42\textwidth}
			\includegraphics[trim={0 0 0 0},clip,height=4.4cm]{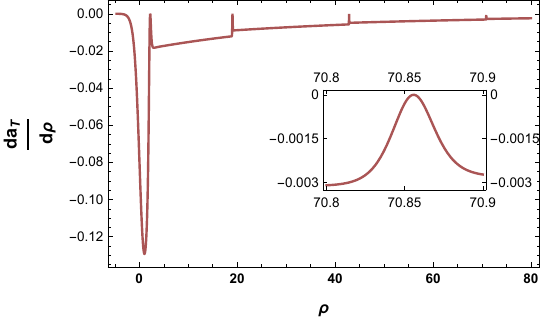}
		\end{subfigure}
		\caption{The thermal $a$-function  and its derivative (in units of $2\pi\ell_P^{-2}$) for RG flows with super-exponential potential $V(\phi)=\frac{1}{2}m^2\phi^2+\frac{1}{10}\exp[\frac{\phi^8}{10}]$ with $\Delta=2$ and $\frac{\phi_0}{T}=6.25$. The Kasner transitions occur where the $a$-function becomes flat or, equivalently, where its derivative gets a steep jump, reaching $\frac{da_T}{d\rho}=0$.}
		\label{thermalafunction}
\end{figure}

In terms of the $\rho$ coordinate, the thermal $a$-function (\ref{aFunctionSch}) reads
 \begin{equation}
a_T(\rho) = \frac{\pi^{d/2}}{\Gamma\left(\frac{d}{2}\right)\ell_P^{d-1}} e^{-(d-1)\chi(\rho)/2}.\label{aFunctionrho}
\end{equation}
Since $a_T(\rho)$ is defined all the way to $\rho\to\infty$ we expect it should be able to diagnose all the cosmological history of the black hole interior. Interestingly, we find sharp features in the $a$-function that could be used to diagnose the Kasner epochs of the black hole interior as well as the transitions between them:
\begin{itemize}
\item First, each Kasner epoch is characterized by a specific dependence of the $a$-function, given by (\ref{aTkasner}). In terms of the $\rho$ coordinate, we find that
\begin{equation}
a_T(\rho)\propto e^{-\frac{(d-1)[d(1+p_t)-2]\rho}{(1-p_t)}}\,,
\end{equation}
possibly up to a constant. Thus, $d  \log a_T / d\rho=(1/a_T)d a_T / d\rho$ is a constant if and only if the geometry is approximately Kasner. This constant can be used to read off the value of $p_t$.
\item Second, for each cosmological bounce the $a$-function displays an instantaneous plateau, implying that 
\begin{equation}
d a_T(\rho) / d\rho=0\,.
\end{equation}
This can be deduced by observing that $d a_T / d\rho \propto d \chi / d\rho\propto d \phi / d\rho$, which follows from (\ref{aFunctionrho}) and (\ref{EE:bouncechiphi}), respectively, and the fact that the latter derivative vanishes instantaneously when there is a Kasner transition (see Fig.~\ref{phiphidotrho}). Cosmological bounces are thus identified as \emph{new} fixed points of the RG flow.
\end{itemize}
The above properties are nicely illustrated in the example shown in Fig.~\ref{thermalafunction}. It is worth noting that these features are not present in the case of standard RG flows with quadratic potential, as explored in the previous sections (see Fig.~\ref{dipafunction}). 

\begin{figure}
     \begin{subfigure}{0.42\textwidth}
			\includegraphics[trim={0 -0.15cm 0 0},clip,height=4.4cm]{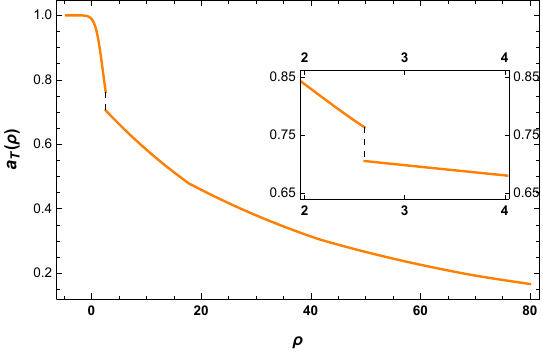}
		\end{subfigure}
		\hspace{1cm}
		\begin{subfigure}{0.42\textwidth}
			\includegraphics[trim={0 0 0 0},clip,height=4.4cm]{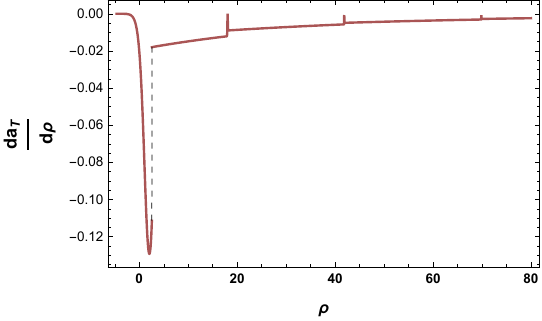}
		\end{subfigure}
		\caption{The thermal $a$-function and its derivative under a shock wave for the same parameters as Fig.~\ref{thermalafunction}. Depending on $E/M$, the time skip can completely bypass one of the cosmological bounces, or even jump over one or multiple Kasner epochs.}
		\label{thermalafunctionshock}
  \end{figure}

We can also study the influence of a shock wave sent at some arbitrary time $t_w$ on the thermal $a$-function. In this case, the interior evolution may abruptly transition between epochs or skip one or multiple epochs altogether. The time at which the skip occurs is controlled by $t_w$, while the strength of the skip is determined by the ratio $E/M$. We illustrate how these behaviors are captured by the $a$-function in Fig.~\ref{thermalafunctionshock}.

\subsection{Butterfly \& entanglement velocities}
For RG flows with quadratic potential, we found that butterfly and entanglement velocities are single-valued functions of the boundary deformation parameter, $\frac{\phi_0}{T^{d-\Delta}}$, displaying a non-monotonic behavior. Moreover, these observables were found to be multi-valued functions of the Kasner exponent $p_t$, implying that they cannot uniquely determine the nature of the singularity or vice-versa. 
In this section, we study these observables for black holes with bouncing interiors. The goal is to investigate whether any distinctive signatures can be observed.

\begin{figure}
		\centering
		\begin{subfigure}{0.45\textwidth}
			\centering
			\includegraphics[width=\textwidth]{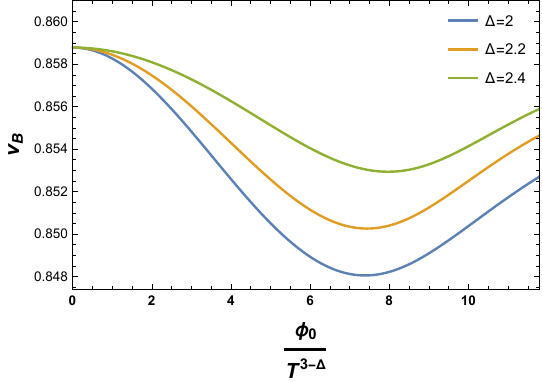}
		\end{subfigure}
		\hfill
		\begin{subfigure}{0.45\textwidth}
			\centering
			\includegraphics[width=\textwidth]{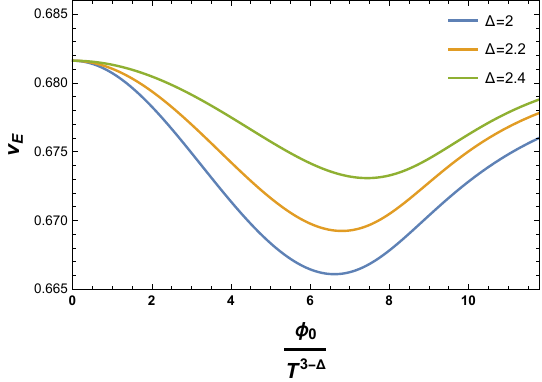}
		\end{subfigure}
		\caption{Butterfly and entanglement velocity as a function of boundary deformation $\phi_0/T^{3-\Delta}$ for RG flows with super-exponential potential $V(\phi)=\frac{1}{2}m^2\phi^2+\frac{1}{10}\exp[\frac{\phi^8}{10}]$ with $\Delta=2$. These observables display the same qualitative behavior as we obtained for RG flows with quadratic potential.}
		\label{velocitiesbounces}
  \end{figure}

Following the same steps outlined in the previous section we find that, in terms of the $\rho$ coordinate,
\begin{equation}
    v_B=\sqrt{{\pi T e^{\rho_h} e^{-\chi_h/2}}}\,,\qquad v_E=\frac{e^{4\rho_h}|f|e^{-\chi}}{e^{4\rho}}\bigg|_{\rho=\rho_{\text{crit}}}\,,
\end{equation}
where $\rho_{\text{crit}}$ is the value of the turning point $\rho_t$ which maximizes $\mathcal{E}^2=-\frac{f(\rho_t)e^{-\chi(\rho_t)}}{e^{{4\rho_t}}}$. As expected, $v_B$ can be written in terms of horizon data, while $v_E$ does generally depend on the interior geometry since $\rho_{\text{crit}}>\rho_h$. In Fig.~\ref{velocitiesbounces} we show the results for $v_B$ and $v_E$ as a function of $\frac{\phi_0}{T^{d-\Delta}}$.
The qualitative nature of the dependence of these observables on the deformation is the same as in the case where the interior has no cosmological bounces. We can understand this for the butterfly velocity, as the (super-)exponential part of the potential only kicks in for values of $\rho$ larger than the horizon, having a minimal effect on the exterior geometry. As for the entanglement velocity, we observe that $\rho_{\text{crit}}$ typically gets stuck at the first of the Kasner epochs and does not really reach the region where the bounces take place. We have plotted  $v_B$ and $v_E$ as a function of $p_t$ corresponding to the first of the Kasner epochs but the end results are qualitatively similar to those in Fig.~\ref{butfig} and Fig.~\ref{ev}, respectively, so we will not show them here. Finally, it is easy to check that the bound (\ref{ratiovevb}) is satisfied for this case as well.

\section{Conclusions and outlook}{\label{discussion}}
In this paper, we studied holographic RG flows at finite temperature perturbed by a shock wave. In particular, we considered RG flows generated by deforming the boundary CFT with a relevant operator, altering the geometry of the black hole interior from AdS-Schwarzschild to a more general Kasner universe near the spacetime singularity. We then introduced null matter in the form of a shock wave into this deformed background and investigated the imprints of various field theory observables on and from the near-horizon and interior dynamics of the black hole. 

Using the out-of-time-order correlators, we found that the scrambling time monotonically increases as the strength of the deformation is increased, while the butterfly velocity exhibits a non-monotonic behavior. These two observables are extracted from a `chaotic' limit of the OTOC and are completely determined from near-horizon data. However, while expressing them in terms of near-singularity data, we learned that they are generically multivalued functions of the Kasner exponent $p_t$, implying that subleading corrections to the Kasner regime are needed to fully specify them. This should not come as a surprise. In holography, we are used to specify the asymptotic values of the bulk fields (non-normalizable modes) as well as the subleading values (normalizable modes) to fully determine a bulk solution. Similarly, the initial value problem in general relativity requires us to specify not only the metric of a spatial slice but also its conjugate momentum (or its extrinsic curvature), which contains information of its derivatives. In that spirit, it would be interesting to understand how subleading corrections to the Kasner regime map to near-horizon data and vice versa for some gravitational solutions of interest. Finally, inspired by the recent results of \cite{Horowitz:2023ury}, we uncovered a novel `hybrid' OTOC that can probe the interior geometry and reach all the way to the singularity for $d\geq3$. Specifically, in a particular limit, we showed that his OTOC includes a term with a distinctive non-analytic dependence on $p_t$, shown in (\ref{OTOTpt}), providing a way to fully determine the Kasner geometry near the singularity. It would be interesting to come up with a bulk representation of such OTOCs in terms of the scattering of quanta in the black hole interior, perhaps following the steps of \cite{Shenker:2014cwa}. Similarly, it would be very insightful to reproduce (\ref{OTOTpt}) from a field theory calculation and to understand precisely how is $p_t$ encoded in the deformed CFT.

Next, we focused on two observables that are more sensitive to the black hole interior: the thermal $a$-function and the entanglement velocity. We showed that, because of the shock wave, the $a$-function undergoes a finite discontinuity, signaling an instantaneous loss of degrees of freedom due to the infalling matter. The discontinuity can be explained by observing that its derivative is proportional to the bulk stress-energy tensor which, in the case of a shock wave, takes the form of a delta function. Thus, after integrating, it generates the advertised discontinuity. In terms of the interior's evolution, we interpret this discontinuity as a `cosmological time
skip,’ which arises as a result of an infinitely boosted length contraction. Further, for matter respecting the null energy condition, this time skip can be shown to always be future-directed, thus, protecting the chronology of the `cosmological interior.' This result follows almost immediately from the monotonicity of the $a$-function. It would be interesting to investigate possible new phenomena at the semi-classical level, in which case the NEC need not be satisfied. Regarding the entanglement velocity, we found a very similar behavior to the butterfly velocity, i.e., a non-monotonic dependence with respect to the strength of the deformation. Even though the entanglement velocity does probe part of the black hole interior, we arrived at the same obstacle when expressing it in terms of near-singularity data, namely, that it is generally a multivalued function. This follows from the fact, that the surfaces that compute this quantity generically get stuck at some final slice and thus do not actually probe the near-singularity region.

Lastly, we repeated our analyses in a model where
the interior geometry undergoes an infinite sequence of bouncing Kasner epochs, introduced originally in \cite{Hartnoll:2022rdv}. In this case, we found that the $a$-function presents very distinctive features that could be used to diagnose the Kasner epochs of the interior as well as the transitions between them. Notably, the $a$-function becomes stationary exactly as each of the cosmological bounces take place, which are thus interpreted as new fixed points in the RG flow. Further, as a result of the `time skip,' when we perturb the system with a shock wave the interior's evolution may now abruptly transition between epochs or even skip one or multiple epochs altogether. Other observables, such as the butterfly velocity or the entanglement velocity, exhibit limited sensitivity to the intricate interior dynamics. This lack of sensitivity is especially anticipated in the case of the butterfly velocity, as it can be exclusively determined from near-horizon data, and this is minimally influenced by the dynamics occurring in the interior. Regarding the entanglement velocity, we find it fails to probe very deep in the interior, resulting in a qualitatively similar behavior to scenarios without cosmological bounces. In the future, it would be interesting to investigate the behavior of these and other observables in other models with exotic cosmological interiors, for example, in the case of holographic superconductors \cite{Hartnoll:2020fhc,Sword:2021pfm,Sword:2022oyg}.

 \acknowledgments

 We would like to thank José Barbón, César Gómez and Karl Landsteiner for interesting discussions and comments on the manuscript. The work of EC is supported by the National Science Foundation under grant number PHY-2112725. AKP and JFP are supported by the ‘Atracci\'on de Talento’ program (Comunidad de Madrid) grant 2020-T1/TIC-20495, by the Spanish Research Agency via grants CEX2020-001007-S and PID2021-123017NB-I00, funded by MCIN/AEI/10.13039/501100011033, and by ERDF A way of making Europe. EC  thanks the Instituto de F\'isica Te\'orica UAM/CSIC, Madrid, for hospitality during the initial stages of this work.
 
\bibliographystyle{JHEP}
\bibliography{main.bib}
\end{document}